\documentclass[a4paper,11pt]{article}

\usepackage[T1]{fontenc}
\usepackage[english]{babel}
\usepackage[utf8]{inputenc}
\usepackage[normalem]{ulem}
\usepackage{comment}
\usepackage{float}

\usepackage[vmargin=2.4cm,hmargin=2.4cm,includeheadfoot]{geometry}
\usepackage[skip=0.25\baselineskip,indent,parfill]{parskip}

\usepackage{color}
\usepackage{subcaption}
\usepackage{tikz}
\usepackage{pgfplots}
\usetikzlibrary{arrows.meta}
\usetikzlibrary{tikzmark}
\usetikzlibrary{shapes.multipart}
\usetikzlibrary{decorations.markings}
\usetikzlibrary{pgfplots.fillbetween}
\pgfplotsset{compat=1.18}

\definecolor{darkblue}{rgb}{0.1,0.1,.7}

\usepackage[colorlinks, linkcolor=darkblue, citecolor=darkblue, urlcolor=darkblue, linktocpage, pagebackref]{hyperref}

\renewcommand*{\backref}[1]{}
\renewcommand*{\backrefalt}[4]{%
  \ifcase #1 %
  \or
    {\scriptsize\,[p.~#2]}%
  \else
    {\scriptsize\,[pp.~#2]}%
  \fi
}

\usepackage[dont-mess-around]{fnpct}

\usepackage[numbers,sort&compress]{natbib}

\makeatletter
\patchcmd\NAT@citexnum{\let\NAT@last@num\NAT@num}{\MakeLinkTarget[cite]{}\Hy@backout{\@citeb\@extra@b@citeb}\let\NAT@last@num\NAT@num}{}{\fail}
\makeatother

\usepackage{booktabs}

\usepackage{amsmath}
\usepackage{amssymb}
\usepackage{mathrsfs}
\usepackage{amsthm}
\usepackage{mathtools}
\usepackage{enumitem}
\usepackage{braket}

\usepackage{multirow}
\usepackage{nicematrix}

\numberwithin{equation}{section}

\let\originalleft\left
    \let\originalright\right
\renewcommand{\left}{\mathopen{}\mathclose\bgroup\originalleft}
    \renewcommand{\right}{\aftergroup\egroup\originalright}

\newcommand{\be}{\begin{equation}} \newcommand{\ee}{\end{equation}}

\renewcommand{\tilde}{\widetilde}

\usepackage{xifthen}
\newcommand{\dif}[1][]{\mathord{\ifthenelse{\isempty{#1}}{\mathrm{d}}{\mathrm{d}^{#1}}}}

\begin{document}

\begin{titlepage}
    \vspace*{1em}
    \begin{center}
        {\LARGE \bfseries Primal Bootstrap for Pion Scattering at Large-$N_c$} \\[3em]
        {\bf
            Sebastiano Bocchia$^{a,b}$, Alessandro Vichi$^{b}$ 
        } \\[2em]
         \( {}^{a} \) {\itshape School of Science and Technology, City St George's, University of London, \\ Northampton Square, London EC1V 0HB, United Kingdom} \\
        \( {}^{b} \) {\itshape Department of Physics, University of Pisa and INFN, \\Largo Pontecorvo 3, I-56127 Pisa, Italy} \\[2em]
        {\small{\texttt{sebastiano.bocchia@city.ac.uk,  alessandro.vichi@unipi.it}}}
    \end{center}
    \vspace{3em}
    \begin{abstract}

We introduce a basis for tree-level meromorphic scattering amplitudes suitable for describing pion scattering in the large-$N_c$ limit. The basis is constructed as linear combinations of Lovelace–Shapiro–like amplitudes with varying Regge slopes and intercepts. The resulting amplitudes satisfy -- by construction -- the fundamental requirements of analyticity, crossing symmetry, and Regge behavior.
We analyze their behavior in specific kinematical regimes, including the high-energy fixed-angle limit. We also show that finite linear combinations of our basis elements need not violate unitarity. Nonetheless, because unitarity is not imposed by construction, we enforce it a posteriori by requiring positivity of the partial-wave decomposition. This condition can be formulated as an optimization problem and solved numerically. The solutions to this primal bootstrap problem yield meromorphic amplitudes that satisfy all the aforementioned constraints. We compare several observables with the bounds obtained from the dual positivity conditions and show that our family of amplitudes spans the full allowed parameter space.
With appropriate modifications, this method can be extended to construct amplitude families for broader applications.
    \end{abstract}
\end{titlepage}

\tableofcontents

\section{Introduction and Summary of Results}

The S-matrix bootstrap represents a powerful non-perturbative approach to theoretical physics, which aims to infer observables from the properties of the system without relying on a specific model. In recent years, the S-matrix bootstrap approach has experienced a remarkable renaissance, especially in the study of quantum chromodynamics (QCD) scattering amplitudes.

This philosophy is closely related to the modern conformal bootstrap, where general principles such as symmetry, crossing symmetry, and unitarity are turned into quantitative constraints on the space of consistent theories \cite{Rattazzi:2008pe} (see \cite{Poland:2018epd} for a review). In the S-matrix context, a particularly powerful realization of this idea is provided by positivity bounds: analyticity, crossing symmetry, polynomial boundedness, and unitarity imply non-trivial constraints on the coefficients of the low-energy effective field theory \cite{Arkani-Hamed:2020blm,Bellazzini:2020cot,Tolley:2020gtv,Caron-Huot:2020cmc}. Most applications of these ideas are naturally ``dual'' in spirit: they derive inequalities that carve out regions of parameter space that cannot arise from a standard ultraviolet completion.

The aim of the present work is complementary. Inspired by \cite{Haring:2023zwu,Veneziano:2017cks}, we develop a primal approach to positivity bounds for pion scattering at large $N_c$. Instead of starting from dispersion relations and deriving exclusion bounds on EFT data, we construct explicit meromorphic amplitudes that satisfy analyticity, crossing symmetry, and Regge behavior by construction, and then impose tree-level unitarity a posteriori through positivity of the partial-wave residues. In this way, we fill the allowed space by producing actual amplitudes, rather than only bounding it from the outside. This has the additional advantage that each point in the primal region comes with a concrete resonance spectrum and a set of pion couplings, allowing us to study the microscopic organization of the extremal solutions.

The elastic scattering of pions provides a particularly compelling laboratory for S-matrix bootstrap techniques. This amplitude has been extensively studied in previous literature, thereby underscoring its significance for this approach \cite{Mandelstam:1958xc,Guerrieri:2018uew,Guerrieri:2020bto,Correia:2020xtr,Albert:2022oes,Albert:2023jtd,Albert:2023seb,Fernandez:2022kzi,Ma:2023vgc,Li:2023qzs}. A particularly significant advantage of studying pion scattering in the context of QCD is that many of the assumptions typically required in bootstrap approaches are satisfied due to the underlying gauge theory structure. 
This makes pion scattering an ideal testing ground for bootstrap methods and provides confidence in the reliability of the results obtained.

To make contact with previous results \cite{Albert:2022oes,Albert:2023jtd,Albert:2023seb,Veneziano:2017cks,Haring:2023zwu}, we study a particularly simplified framework of QCD: the large-$N_c$ limit with massless quarks. The large-$N_c$ limit of QCD \cite{HOOFT1974461,WITTEN197957} provides theoretical control over the non-perturbative dynamics of the strong interactions. In this limit, meson-meson scattering amplitudes are suppressed by powers of $1/N_c$, and the leading contributions correspond to tree-level exchanges of physical mesons. This simplification makes the large-$N_c$ limit particularly appealing for an S-matrix bootstrap analysis while still capturing essential phenomenological aspects of QCD, such as confinement, chiral symmetry breaking, and the emergence of an infinite tower of narrow resonances.

In this work, we introduce a generalized family of amplitudes that preserves the desirable analytic and asymptotic properties of Regge theory. Starting from the Lovelace--Shapiro amplitude \eqref{Lovelace-Shapiro model}, originally proposed as a variation of the Veneziano model to describe tree-level elastic scattering of massless pions in the large-$N_c$ limit \cite{Lovelace:1968kjy,Shapiro:1969km}, we extend the construction to a broader class of meromorphic amplitudes. The generalized ansatz is built from Lovelace--Shapiro--like building blocks with different slopes and intercepts, chosen so that crossing symmetry, analyticity, and Regge behavior are manifest, while leaving the particle content/analytic structure unchanged. This is done in Section~\ref{A generalized Lovelace-Shapiro Model}, where we introduce our Generalized Lovelace-Shapiro (GLS) construction and analyze its analytic properties in detail. The asymptotic behavior in both the Regge limit and the fixed-angle high-energy regime is studied, revealing non-trivial dynamics that depend on the detailed structure of the model parameters. We also discuss the limitations of our analytic approach.

The remaining condition, tree-level unitarity, is imposed numerically through positivity of the partial-wave residues. Section~\ref{Bootstrap Algorithm} describes the numerical S-matrix bootstrap algorithm employed to impose unitarity constraints and determine the allowed parameter space of our models.

    \begin{figure}[h!]
        \centering
        \includegraphics[width=0.8\textwidth]{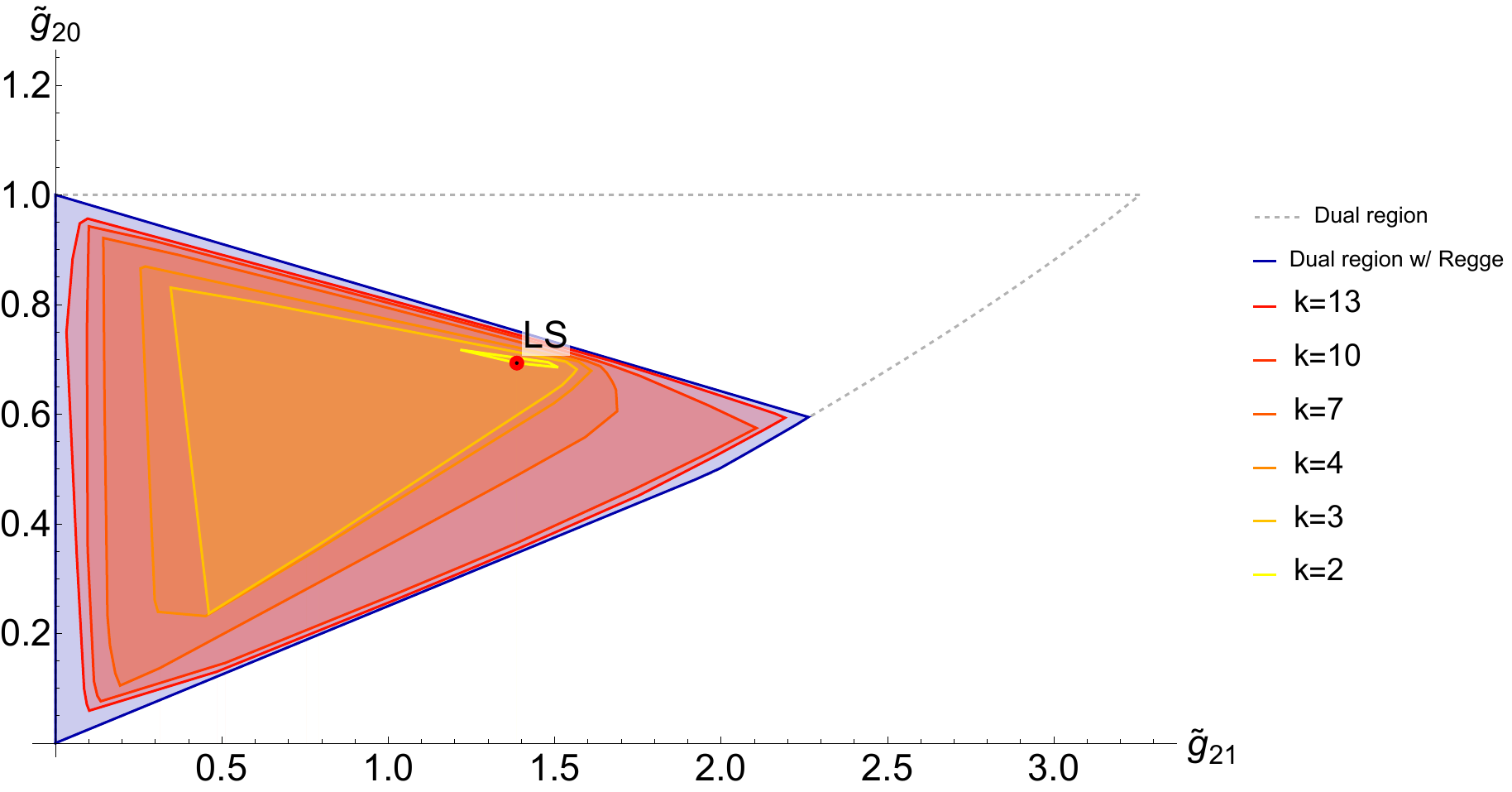}
        \caption{Ruled-in regions in the space of two low energy coefficients $\tilde{g}_{21},\tilde{g}_{20}$ covered by the GLS ansatz (regions yellow to red correspond to GLS amplitudes with increasing number of parameters; red dot is the LS amplitude). The dark blue region is the region allowed by dual bootstrap bounds assuming that for each spin $J$ the resonances have masses $M^2> f(J)$, with $f(J)$ a linear Regge trajectory of given slope. See Section~\ref{Results} for details. The dashed line reports the dual bootstrap bound with no assumptions \cite{Albert:2022oes}}
        \label{fig:LS-intro}
    \end{figure}

The goal of this work is not only to present this family of GLS amplitudes, but also to explore how much of the space of admissible pion amplitudes it can cover. In particular, we aim to understand the extent to which this basis is general, how it compares with standard positivity bounds, and in which regimes it fails to reproduce specific low-energy or high-energy features. In Section~\ref{Results} we present our numerical results on low-energy effective field theory coefficients. As shown in Figure~\ref{fig:LS-intro}, ~\ref{fig:LS ideal}, and \ref{fig:g20g30}, most of the  region allowed by dual bootstrap methods \cite{Albert:2022oes,Fernandez:2022kzi} can be ruled-in by the GLS amplitudes, thus showing that the ansatz is general enough to reproduce the Wilson coefficients. \\
We also consider amplitudes that maximize on-shell coefficients, such as the coupling of pions to the $f_2$ meson. The ruled-in region is shown in Figure~\ref{fig:gpipif2} and we inspect the corresponding extremal spectra in Appendix~\ref{Spectra of the Extremal Solutions}. 

Finally, in Section~\ref{Fixed angle scattering} we explore how to impose the correct fixed angle limit predicted for Pion scattering. We show that this constraint has a marginal impact on the ruled-in region for low energy coefficients and we propose alternative observables that can potentially be more sensitive to this limit.  
We conclude in Section~\ref{Conclusions} with a discussion of our findings and future directions.

\section{Set-up}\label{Set-up}
To fully exploit the potential of the S-matrix bootstrap approach, it is essential to incorporate, whether by construction or through numerical methods, all properties expected from our understanding of the underlying physical phenomena.
These assumptions are motivated by the requirements of Regge theory and large-$N_c$ QCD in the chiral limit. Accordingly, our amplitudes are constructed to remain consistent with these principles. This problem has been extensively studied in the existing literature, against which we aim to compare our results. Consequently, the assumptions in this work are closely aligned with those in \cite{Albert:2022oes,Albert:2023jtd,Albert:2023seb,Veneziano:2017cks,Haring:2023zwu}.

\paragraph{\texorpdfstring{Large-$N_c$ QCD in the chiral limit}
                          {Large-Nc QCD in the chiral limit}}
We study QCD in the limit of a large number of colours $N_c$. We adopt the standard assumption that the theory remains confining at large-$N_c$, such that the asymptotic states consist of colour-singlet glueballs, mesons, and heavy baryons. The scaling of various observables with $N_c$ is well-established \cite{WITTEN197957}: meson and glueball masses remain finite, while baryon masses scale as $N_c$. In the exact chiral limit, the quark masses are set to zero and the global flavour symmetry $SU(N_f)_L \times SU(N_f)_R$ is spontaneously broken to the diagonal subgroup $SU(N_f)_V$. The resulting Goldstone bosons are identified with pions, which therefore have derivative interactions as required by the non-linear realization of chiral symmetry \cite{Weinberg:1996kr}. 
In this work, we consider the elastic scattering amplitude between two pions $M_{ab}^{cd}(s,t)$. In the chiral limit, the pion–pion elastic scattering amplitude should exhibit the so-called Adler zero \cite{PhysRev.137.B1022, Weinberg:1965nx}: $
    M_{ab}^{cd}(0,0)=0\,
$ due to the derivative interaction nature of exact Goldstone couplings.

\paragraph{Crossing-Symmetry}

Crossing symmetry for scalar charged particles implies that it is possible to exchange the incoming and outgoing momenta in the amplitude. For a $2 \to 2 $ scattering, this implies that $M(s,t)=M(t,s)$, where we have introduced the standard Mandelstam variables $s,t$.\footnote{In our notation $s=(p_1+p_2)^2, t=(p_1-p_3)^2, u=(p_1-p_4)^2 $ and $s+t+u=0$.}

The amplitude does not exhibit full crossing symmetry, because the external quarks fix the order of the asymptotic states, preventing us from exchanging the two outgoing particles \cite{Albert:2022oes}.

\paragraph{Isospin Decomposition}
In the massless limit for the quarks, isospin symmetry becomes exact and the amplitude can be decomposed into its isospin channels. With $N_f=2$, the decomposition takes the form (see \cite{Albert:2022oes} for the generalisation to $N_f$ flavours, with analogous results):
\begin{equation}
    M_{ab}^{cd}(s,t)=\sum_{I=0}^2M_I(s,t)\mathbb{P}_I
\end{equation}
where $\mathbb{P}_I$ are the isospin projection operators and $M_I(s,t)$ are the amplitudes in channels with total isospin $I$. The final result is:
\begin{equation}\begin{split}
    M_0(s,t)=&\,3M(s,t)+3M(s,u)-M(t,u)\\
    M_1(s,t)=&\,M(s,t)-M(s,u)\\
    M_2(s,t)=&\,M(t,u)
    \label{Isospin decomposition}
\end{split}\end{equation}

\paragraph{Analytic Properties}
The analytic properties of scattering amplitudes represent a well-established area of investigation within theoretical physics, having been the subject of extensive study in the existing literature. The scattering amplitude is usually regarded as the boundary value of a complex function:
\begin{equation}
    M(s,t):= \lim_{\epsilon \to 0^+} M(s+i \epsilon,t) \, .
\end{equation}
This result originates in perturbation theory and is typically extended as a principle in non-perturbative QFT.
This latter assumption ensures real analyticity, relying only on the CPT theorem \cite{Olive1962}:
\begin{equation}
    M(s^*,t^*)=M^*(s,t) \, .
\end{equation}
Within the $S$-matrix bootstrap framework, the amplitude is usually assumed to be analytic over the entire complex plane in both variables. This extended analyticity property is much stronger than what can be rigorously proven in full QCD, where analyticity is established only in limited kinematic regions. However, in the large-$N_c$ limit, this property arises naturally from the tree-level nature of the leading diagrams.

Fixing the angular variable in its physical domain $t<0$, the amplitude is analytic in the entire $s$-plane, except for simple poles on the real axis due to the exchange of massive mesons in the $s$- and $u$-channels. In addition, mesons with isospin $I=2$ cannot be exchanged, and diagrams where the initial and final states are connected solely by gluon propagators are suppressed by the OZI rule \cite{OKUBO1963165,Zweig:570209,10.1143/PTPS.37.21}. Therefore, $M_2(s,t)$ must be analytic for fixed $t<0$ along the real $s$-axis. One finds that $M(s,t)$ has poles only for $s \geq 0$, while no poles appear in the $u$-channel (note, however, that the amplitude $M(s,-s-u)$, for fixed $u$, should be regarded as independent and may display crossed-channel poles for $s<0$) \cite{Albert:2022oes}.\\

\paragraph{Unitarity and Partial-Wave Expansion}
We can rewrite an analytic scattering amplitude as a polynomial expansion; however, we can always choose a more convenient complete basis. For elastic scattering between scalars, the basis of the Legendre polynomials corresponds to decomposing the amplitude in terms of the exchanged angular momentum:
\begin{equation}
    M(s,t) =16 \pi \sum_J (2 J+1) A_J(s) P_J(\cos \theta)
    \label{Partial wave decomposition}
\end{equation}
At tree level, the lower unitarity bound becomes $ \rho_J(s) \geq 0 $. Also,
at tree level, the imaginary part of the amplitude is non-zero only in the presence of particle exchanges. In particular, from the partial wave decomposition \eqref{Partial wave decomposition}, if a particle is exchanged in a spin-$J$ channel, the amplitude receives a contribution of the form:
\begin{equation}
    M(s,t) \sim -\dfrac{g^2}{2}\dfrac{1}{s-m^2+i\epsilon}P_J\Big{(}1+\frac{2t}{s}\Big{)}=-\dfrac{g^2}{2}\Big{(}\mathcal{P} \Big{(} \dfrac{1}{s-m^2}\Big{)}-i\pi \delta(s-m^2) \Big{)}P_J\Big{(}1+\frac{2t}{s}\Big{)}
    \label{Discontinuity from particle exchange}
\end{equation}
and therefore the unitarity condition becomes $g^2>0$. 
This bound is equivalent to requiring that the couplings entering the effective Lagrangian are real.\\
For resonances of the form given in \eqref{Discontinuity from particle exchange}, the positivity condition on the coefficients of the partial wave expansion reduces to a negativity condition on the coefficients of the residues.

\paragraph{Regge Limit}
A further common assumption, introduced in the seminal work \cite{Regge:1959mz}, requires that the partial waves defined in \eqref{Partial wave decomposition} be analytic functions not only of the Mandelstam variable $s$ (for fixed $J$), but also of the complex angular momentum $J$ (for fixed $s$). This assumption is known as the extended analyticity of the second kind \cite{Collins1984}.\\
Analyticity, unitarity and crossing-symmetry, plus the assumption of extended analyticity of the second kind, allow one to relate the high-energy behavior of the amplitude with the physical spectrum of the theory. More specifically, using the $s-t$ crossing symmetry, it is possible to show that a polynomial asymptotic behavior:
\begin{equation}
    M(s,t) \sim s^{\alpha(t)} \qquad (s \rightarrow \infty, \text{fixed }t)\,,
    \label{regge limit formula}
\end{equation}
generates a pole in the $J$-complex plane of the (crossed) partial waves $A_J(t)$ at $J=\alpha(t)$:
\begin{equation}
    A_J(t) \sim -\dfrac{1}{\alpha(t)-J}.
    \label{pole in the complex angular momentum}
\end{equation}
Conversely, each pole in the complex angular momentum plane \eqref{pole in the complex angular momentum} generates an asymptotic behavior of the kind \eqref{regge limit formula}.\\
Whenever $\alpha(t_r) = J$ assumes an integer value, the corresponding resonance becomes a physical resonance as it enters the sum in \eqref{Partial wave decomposition}, and may be interpreted as a bound state of spin $J$ and mass squared $m_r^2 = \Re \{t_r\}$. Analogously, the imaginary part of the pole $\Im\{ t_r\}= m_r \Gamma$ can be interpreted as the decay width of the resonance \cite{Collins:1977jy}. The function $\alpha(t)$ is known as the Regge trajectory. In the limit $t\rightarrow0$, the Regge trajectory is approximated by a linear function $\alpha(t) \simeq \alpha_0 + \alpha' t$, where $\alpha_0$ is the  Regge intercept and $\alpha'$ is the Regge slope.\\
QCD shows good phenomenological agreement with Regge theory; in particular, the spectrum is organised into Regge trajectories. At finite-$N_c$, the high-energy behavior (at fixed $t$) is controlled by the Regge trajectory of the pomeron, which corresponds to a glueball state. As we have already pointed out, glueball–meson interactions are suppressed at leading order, and we should therefore not expect its presence in this limit. The dominant trajectory after the pomeron is given by the $\rho$ and the $P'$ trajectories, associated respectively with the $\rho$ meson and the $f_2(1270)$ meson; these trajectories share approximately the same Regge intercept $\alpha(0) \approx 0.52$, and we therefore interpret them as the same trajectory.\\
As argued in \cite{Veneziano:2017cks}, it is not clear whether QCD in the large-$N_c$ limit should obey Regge behavior, since it is not trivial that the large-$N_c$ limit and the Regge limit commute. In our model, we will treat Regge behavior and the presence of linear Regge trajectories as assumptions, motivated by phenomenological observations. The requirement of Regge behavior implies that the residues in $s$ are polynomial in $t$, bounding the spin of the possible exchanged particles.\\

In \cite{Albert:2022oes}, given that the intercept of the leading trajectory is approximately:
\begin{equation}
    \alpha(0) \approx 0.52
\end{equation}
and the Regge trajectory is a monotonic function, it has been assumed:
\begin{equation}
    \lim_{|s| \to \infty}\dfrac{M(s,t)}{s}=0
    \label{Regge behavior requirement ansatz}
\end{equation}
for fixed, physical values (i.e. negative) of $t$. This assumption has been used to derive dispersion relations. We will not make direct use of dispersion relations, but we will impose this bound to ensure that our results can be compared with those of \cite{Albert:2022oes}. As we will see, this will be the upper bound that our model is capable of saturating.

\section{A Generalized Lovelace-Shapiro Model}\label{A generalized Lovelace-Shapiro Model}
A remarkable model that satisfies the  properties discussed in the previous section is  the Lovelace–Shapiro (LS) amplitude \cite{Lovelace:1968kjy,Shapiro:1969km}:\footnote{Generically the LS amplitude can be multiplied by an overall positive coupling $g^2$.}
\begin{equation}
M_{LS}(s,t)= \dfrac{\Gamma(1-\alpha(s))\Gamma(1-\alpha(t))}{\Gamma(1-\alpha(s)-\alpha(t))}
\label{Lovelace-Shapiro model}
\end{equation}
This model is governed by its (linear) Regge trajectory $\alpha(s)$, which encodes the information about the meson spectrum and the high-energy behavior of the amplitude:
\begin{equation}
\alpha(s):=\alpha_0+\alpha' s
\end{equation}
The usual choice for the parameters $\alpha'$ and $\alpha_0$ is to normalise the mass of the $\rho$ meson to 1, and to impose the Adler zero requirement, setting $\alpha_0=\alpha'=\tfrac{1}{2}$. Alternatively, the parameters $\alpha_0,\alpha'$ can be related to the masses of the lightest spin 1 and spin 2 resonances (which we denote $\rho$ meson and $f_2$ meson):\footnote{The value of the physical masses is not consistent with the Adler zero requirement, but instead gives $\alpha_0 \simeq 0.42$, $m_\rho^2 \, \alpha'=1-\alpha_0 \simeq 0.58$.}
\begin{equation}\begin{split}
    \alpha'=\dfrac{1}{m^2_{f_2}- m^2_\rho}\,,\qquad
    \alpha_0=1-\dfrac{m^2_\rho}{m^2_{f_2}- m^2_\rho}\,.
    \label{Slope and intercept as a function of the dominant meson}
\end{split}\end{equation}
The amplitude has indeed poles whenever $\alpha(s)$ assumes a positive integer value $N=1,2,\ldots$. The corresponding residue is polynomial in $t$ and, once decomposed in partial waves, it contains resonances with spin equal to $0,1,\ldots ,N$. 
The resonances at level $J$, with spin $J$, belong to the leading Regge trajectory and have mass
\begin{equation}
    m^2(J) = (m_{f_2}^2-m^2_\rho )(J-2) + m_{f_2}^2
    \label{eq:leading-Regge-masses}
\end{equation}
The resonances with smaller spin are organized into an infinite series of daughter trajectories, which are parallel to the leading trajectory.

\subsection{Generalized Lovelace-Shapiro (GLS) amplitude}
The basic idea of this work is to construct a basis for meromorphic scattering amplitudes that satisfies most of the requirements listed in Section~\ref{Set-up} -- in particular crossing, analyticity and Regge behavior -- by construction. The only missing ingredient --  unitarity -- will then be imposed as a constraint on the parameters of the ansatz. 
Hence, we consider linear combinations of the LS amplitudes with varying intercepts and slopes.  This basis generalizes the ansatz introduced in \cite{Veneziano:2017cks}. The resulting ansatz is:
\begin{align}\label{Ansatz}
M(s,t)=&\sum_{n}\sum_{k} c_{n,k}\, A_{n,k}(s,t)\nonumber\\
A_{n,k}(s,t)=&\dfrac{\Gamma(1-\alpha_{n,k}(s))\Gamma(1-\alpha_{n,k}(t))}{\Gamma(1-\alpha_{n,k}(s)-\alpha_{n,k}(t))}\\
\alpha_{n,k}(s):=& \dfrac{(k-n)-1+\alpha_0}{k}+\dfrac{1}{k} \, \alpha' \, s\nonumber
\end{align}

We have chosen the form of $\alpha_{n,k}$ such that: 
\begin{enumerate}
\item the spectrum of resonances is the same as in the original LS amplitude ($k=1, n=0$), otherwise unitarity would have required that all coefficients of the linear combination were positive, $c_{n,k} \geq0$. In our case instead, only certain linear combinations of coefficients --  corresponding to the partial waves -- must be positive. We will come back to this in the next section.
\item to avoid poles on the negative $s$ axis, we only consider the sum over $n > \alpha_0-1$.
\item the Regge behavior is controlled by the leading Regge trajectory of the original LS amplitude. Each term in the sum has a leading trajectory whose spin-1 and spin-2 masses are
\begin{equation}
    m^2_{1}(n,k)=\tfrac{n+1-\alpha_0}{\alpha'},\qquad m^2_{2}(n,k)=m^2_{1}(n,k)+\tfrac{k}{\alpha'}\,.
\end{equation}
It would also be possible to consider negative intercepts, but for the moment we restrict our sum to positive (non-vanishing) ones. This requirement is equivalent to imposing a lower bound on the lightest spin-2 particle, requiring $m^2_{2}(n,k)>m^2_{1}(n,k)$. 
\end{enumerate}

\begin{figure}[H]
    \centering
    \includegraphics[width=0.75\linewidth]{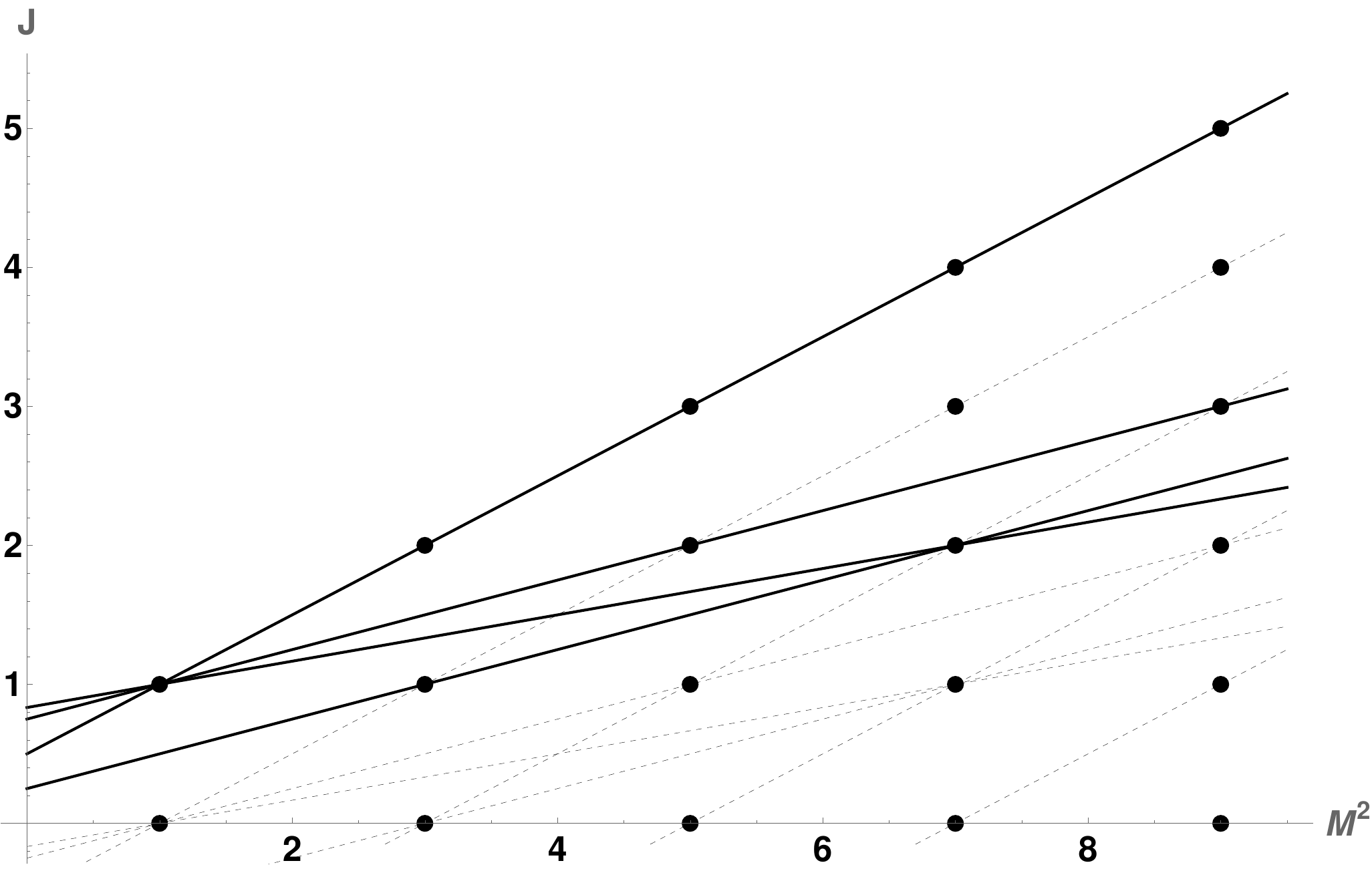}
    \caption{Structure of the Regge poles, for a slope $\alpha'=1/2$. The thick lines represent the leading trajectories of some amplitudes, the dashed lines their corresponding daughter trajectories.}
    \label{fig:regge poles ansatz}
\end{figure}

The most general spectrum of the theory is presented in figure \ref{fig:regge poles ansatz}.\\

 Putting together all the bounds on the indices in the sum, we obtain $\alpha_0-1 < n < \alpha_0-1+k$, which is equivalent to fixing all the values of the Regge intercepts between 0 and 1.\\

As pointed out in \cite{Fernandez:2022kzi}, adding (or removing) spin-0 exchanges to our amplitudes does not affect the analytic properties we have required. In particular, we do not violate the asymptotic behavior predicted by Regge theory. Thus, we consider a more general version of our ansatz:
\begin{equation}\label{Ansatz-scalars}
    \tilde M(s,t) = M(s,t)-\sum_{j}g^2_{\pi \pi m_j} \Bigg(\dfrac{m_j^2}{m_j^2-s}+ \dfrac{m_j^2}{m_j^2-t}\Bigg)
\end{equation}
It would also have been possible to add spin-1 exchanges, which saturate the bounds as explained in \cite{Fernandez:2022kzi}. However, this has not been necessary to obtain the optimal bounds for our ansatz.

\subsection{Regge Limit} \label{Regge Limit}
It is interesting to study analytically specific kinematic regimes of this ansatz. In this section and the following one, we will study specific kinematical regimes using the method developed in \cite{Veneziano:2017cks}.\\
We can distinguish two situations depending on how we fix the parameters $c_{n,k}$. The first situation corresponds to truncating the sum over $k$ or fixing a characteristic scale in $k$ (in certain models, such as the one studied in \cite{Haring:2023zwu}, this choice is problematic; on the contrary, we show in Appendix~\ref{Analytic-Study of Unitarity} that this is not the case here). The second corresponds to having an infinite series. This situation is the most general, as the first one is just a special case of this one, where we have chosen $c_{n,k>k_{max}}=0$.\\
In the Regge limit, we have a sum of Regge behaviors:
\begin{equation}
    \dfrac{M(s,t)}{g^2}=\sum_{k=1}^{k_{\max}} \sum_{n >\alpha_0-1}^{\alpha_0-1+k} c_{n,k}\, \Gamma(1-\alpha_{n,k}(t))\Bigg(-\frac{\alpha'}{k}s\Bigg)^{\alpha_{n,k}(t)}
    \label{Regge limit 1}
\end{equation}
If $k_{max}$ is finite, the leading trajectory will be the highest one, for each fixed $t$:
\begin{equation}
\lim_{\substack{s \to \infty \\ t \, \text{fixed}}}
M(s,t) = s^{\max\{\alpha(t)\}}.
\end{equation}
Note that, for $\alpha' t \geq 1-\alpha_0$, this always corresponds to the trajectory of the original Lovelace-Shapiro.\\
We must be more careful in considering the Regge limit if the sum is unbounded. The explicit computation is in Appendix \ref{Regge Limit appendix}, here we just show the main results.\\
For $\alpha' t > 1-\alpha_0$, the leading behavior of the amplitude is controlled by the Regge trajectory $\alpha_L=\alpha_0+\alpha' \, t$ of the Lovelace–Shapiro model.\\
To evaluate the other regime $\alpha' t < 1-\alpha_0$, we necessarily need some assumption on the asymptotic behavior of the coefficients $c_{0,k}$. In particular, if we assume that they are suppressed with a power-like behavior:
\begin{equation}
    c_{0,k} \sim \dfrac{1}{k^{\beta+1}}
    \label{Asymptotic behavior coefficients}
\end{equation}
with $\beta>0$. Then, the leading behavior in the Regge limit is given by:
\begin{equation}
    M(s,t) \sim s 
    \label{Regge negative t}
\end{equation}

\subsection{Unitarity}

The GLS amplitude satisfies most of the assumptions listed in Section~\ref{Set-up} by construction. Unitarity and the Adler zero remain to be enforced.\\
Unitarity corresponds to the positivity bounds on the imaginary part of the partial waves and must be realized by properly choosing the coefficients in \eqref{Ansatz}. 

Consider a generic term of the GLS amplitude:
\begin{equation}
    M_{LS}(s,t)=\dfrac{\Gamma(1-\alpha(s))\Gamma(1-\alpha(t))}{\Gamma(1-\alpha(s)-\alpha(t))}
\end{equation}
with Regge trajectory:
\begin{equation}
    \alpha(s) := \alpha_0+\alpha' s.
\end{equation}
This amplitude has a pole whenever $s=\frac{N+1-\alpha_0}{\alpha'}$, for each positive integer $N$. The corresponding residue is
\begin{equation}
    \mathrm{Res}\Big[M_{LS}(s,t)\Big|s=\dfrac{N+1-\alpha_0}{\alpha'}\Big]=\dfrac{1}{\alpha'}\dfrac{(-1)^N}{N!}\dfrac{\Gamma(1-\alpha(t))}{\Gamma(-N-\alpha(t))}.
    \label{Residue LS}
\end{equation}
Remarkably, the residues are polynomial in $t$ -- i.e. in $\cos (\theta)$ -- and can be easily expanded in Legendre polynomials $P_J(x)$ using the relation
\begin{equation}
    x^{m}=
\sum_{\substack{J=0 \\ m-J\ \mathrm{even}}}^{m}
(2J+1)\,
\frac{m!}{(m-J)!!\,(m+J+1)!!}\,
P_J(x)
\end{equation}

Going back to the ansatz \eqref{Ansatz}, we can simply substitute  $\alpha_{n,k}(s)$ in \eqref{Residue LS} and obtain the total partial waves. For each resonance at level $N$ and spin $J$ we then obtain a constraint on the positivity of the corresponding partial wave of the form
\begin{equation}\label{SDP-conditions}
    \sum_{n,k} a_{n,k}(N,J) \, c_{n,k} \geq 0 \qquad \text{for all }\, J,N \in \mathbb N, \,\, J\leq N\,.
\end{equation}
Since the coefficients $a_{n,k}(N,J)$ are known, the above infinite set of equations represents  constraints on the parameters $c_{n,k}$.

\subsection{Low-energy expansion}\label{Effective Field Theory results}
Finally, we can also explore the low-energy behavior of the GLS amplitude. Consider the Taylor expansion of the amplitude in the $s,t$ variables:
\begin{equation}
    M(s,t):=\sum_{n=1}^\infty\sum_{l=0}^{n/2}g_{n,l}(s^{n-l} t^{l}+s^{l} t^{n-l})
    \label{Rastelli amplitude}
\end{equation}
The coefficients $g_{n,l}$ can be associated to contact term interactions in the low-energy Effective Field Theory description of pions. Matching the above expression with the ansatz \eqref{Ansatz}, one can express each coefficient as a linear combination of the parameters $c_{n,k}$. For instance, in our model:
\begin{align}\label{eq:low-energy}
  g_{1,0}&=- \sum_{n,k} c_{n,k} \frac{
  2^{\frac{-1 + k - 2n}{k}} \sqrt{\pi}\,
  \Gamma\!\left(\frac{1 + 2n}{2k}\right)
  \left[
    \Psi\!\left(\frac{1 + 2n}{2k}\right)
    -
    \Psi\!\left(\frac{1 - k + 2n}{k}\right)
  \right]
}{
  k\, \Gamma\!\left(\frac{1 - k + 2n}{2k}\right)
}
\\
  g_{2,0}&=\sum_{n,k} c_{n,k}  \frac{
  2^{-\frac{1 + k + 2n}{k}} \sqrt{\pi}\,
  \Gamma\!\left(\frac{1 + 2n}{2k}\right)
  \left[
    \left(
      \Psi\!\left(\frac{1 + 2n}{2k}\right)
      -
      \Psi\!\left(\frac{1 - k + 2n}{k}\right)
    \right)^{2}
    +
    \Psi^{(1)}\!\left(\frac{1 + 2n}{2k}\right)
    -
    \Psi^{(1)}\!\left(\frac{1 - k + 2n}{k}\right)
  \right]
}{
  k^{2}\, \Gamma\!\left(\frac{1 - k + 2n}{2k}\right)
}
\\
g_{2,1}&=\sum_{n,k} c_{n,k} \frac{
2^{-\frac{1 + k + 2n}{k}} \sqrt{\pi}\,
\Gamma\!\left(\frac{1 + 2n}{2k}\right)
\left[
\left(
\Psi\!\left(\frac{1 + 2n}{2k}\right)
-
\Psi\!\left(\frac{1 - k + 2n}{k}\right)
\right)^{2}
-
\Psi^{(1)}\!\left(\frac{1 - k + 2n}{k}\right)
\right]
}{
k^{2}\,
\Gamma\!\left(\frac{1 - k + 2n}{2k}\right)
}
\end{align}
where $\Psi $ and $\Psi^{(1)} $ are the polygamma functions of degree 0 and 1, respectively. Similar expressions hold for all coefficients. 
Since we can always rescale the amplitude to set one of the coefficients to one, in what follows it will be convenient to consider the normalization independent combinations 
\begin{equation}\label{eq:g2g2p}
    \tilde{g}_2=\dfrac{g_{2,0}}{g_{1,0}}   \,, \qquad \tilde{g}_2'=\dfrac{g_{2,1}}{g_{1,0}} \,.
\end{equation}
Also, we set the combination
\begin{align}
g_{0,0}=\sum_{n,k} c_{n,k}\,\dfrac{\Gamma\Big{(}1-\dfrac{(k-n)-1+\alpha_0}{k}\Big{)}^2}{\Gamma\Big{(}1-2\dfrac{(k-n)-1+\alpha_0}{k}\Big{)}}
\end{align}
to zero, either as a further constraint or by simply considering the possibility of removing it by hand with a constant term $-\lambda$ in the amplitude, since this term does not violate any of the assumptions previously made.

\section{A Primal Bootstrap Algorithm} \label{Bootstrap Algorithm}

The strategy to construct valid amplitudes follows the approach of \cite{Haring:2023zwu} and mimics the non-perturbative primal S-matrix bootstrap. In both cases, one starts from a general ansatz for a scattering amplitude (or S-matrix) -- which satisfies by construction certain assumptions -- and imposes the remaining ones as constraints.

In our case, then, given the ansatz \eqref{Ansatz}, subject to the positivity constraints \eqref{SDP-conditions}, we can explore the space of meromorphic scattering amplitudes that we can construct by varying the values of the coefficients $c_{n,k}$. 

As a first application of the method, we focus on two key observables: the quadratic coefficients in the low-energy expansion, previously studied in \cite{Albert:2022oes,Haring:2023zwu,Fernandez:2022kzi}, and the three-point coupling between the pions and the $f_2$ meson \cite{Albert:2023seb}.\\
Let us describe in detail the algorithm to explore the allowed values of one low-energy coefficient, say $g_{2,0}$. Other applications work similarly.
\begin{enumerate}
    \item We first truncate the ansatz~\eqref{Ansatz} to $k\leq k_\text{max}$
    \item We maximize/minimize the linear combination 
    \begin{equation}
        g_{2,0} = \sum_{k=1}^{k_\text{max}}\sum_{n=\alpha_0-1}^{\alpha_0-1+k} c_{n,k} \,\Psi_{n,k}
    \end{equation}
    subject to the  normalization condition
    \begin{equation}
        1 = g_{1,0} = \sum_{k=1}^{k_\text{max}}\sum_{n=\alpha_0-1}^{\alpha_0-1+k} c_{n,k} \,\tilde\Psi_{n,k}
    \end{equation} and the positivity constraints
    \begin{equation}
        \sum_{k=1}^{k_\text{max}}\sum_{n=\alpha_0-1}^{\alpha_0-1+k} a_{n,k}(N,J) \, c_{n,k} \geq 0 \qquad \text{for }\,  J\leq N \leq  N_\text{max}\,.
    \end{equation}
     where the numerical coefficients $\Psi_{n,k}$ and $\tilde \Psi_{n,k}$ are given in \eqref{eq:low-energy}.
    In the above expression $N_\text{max}$ is a large integer number.
    \item We increase $N_\text{max}$ until the result of the optimization is stable.
    \item The outcome of the optimization procedure is the value of $g_{2,0}$, together with a set of coefficients $c_{n,k}$. Plugging these values in the ansatz allows us to reconstruct the full amplitude and check that all partial waves (for $N>N_\text{max}$) are still positive.
    \item We then increase $k_\text{max}$ and study the convergence properties. 
\end{enumerate}

Since this is a primal algorithm, we are not carving out the space of allowed amplitudes by placing rigorous bounds, but rather we are filling in the space by constructing explicit amplitudes that obey all constraints. As such, any dual bound -- with the same set of assumptions -- should contain our solutions. Moreover, as $k_\text{max}$ increases, the space of amplitudes grows. In the example discussed above, this means that the maximal value of $g_{2,0}$ can only increase as $k_\text{max}$ increases.

Finally, given that we start from an explicit ansatz it is possible to impose further restrictions. For instance one could impose the existence of the Adler's zero by requiring $g_{0,0}=0$, or one could maximize certain observables and impose specific values for the low-energy coefficients, or specific relations among them, along the lines of \cite{Albert:2022oes,Albert:2023jtd,Albert:2023seb,McPeak:2023wmq}

\section{Results}\label{Results}

\subsection{Low energy coefficients}

\begin{figure}[H]
    \centering
        \begin{subfigure}[b]{0.45\linewidth}
        \centering
        \includegraphics[width=\linewidth]{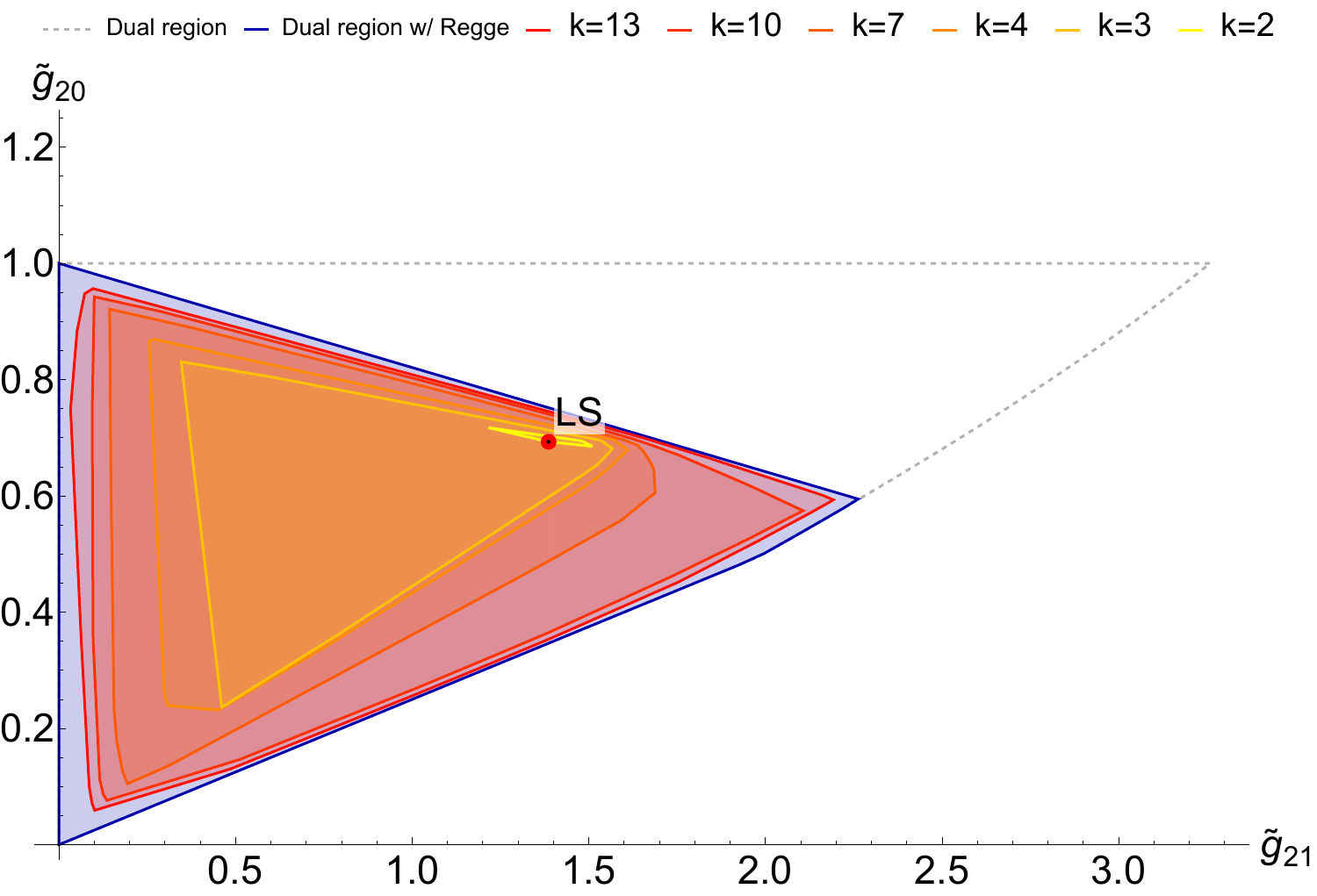}
        \caption{}
        \label{fig:LS no spin zero}
    \end{subfigure}
    \hfill
    \begin{subfigure}[b]{0.45\linewidth}
        \centering
        \includegraphics[width=\linewidth]{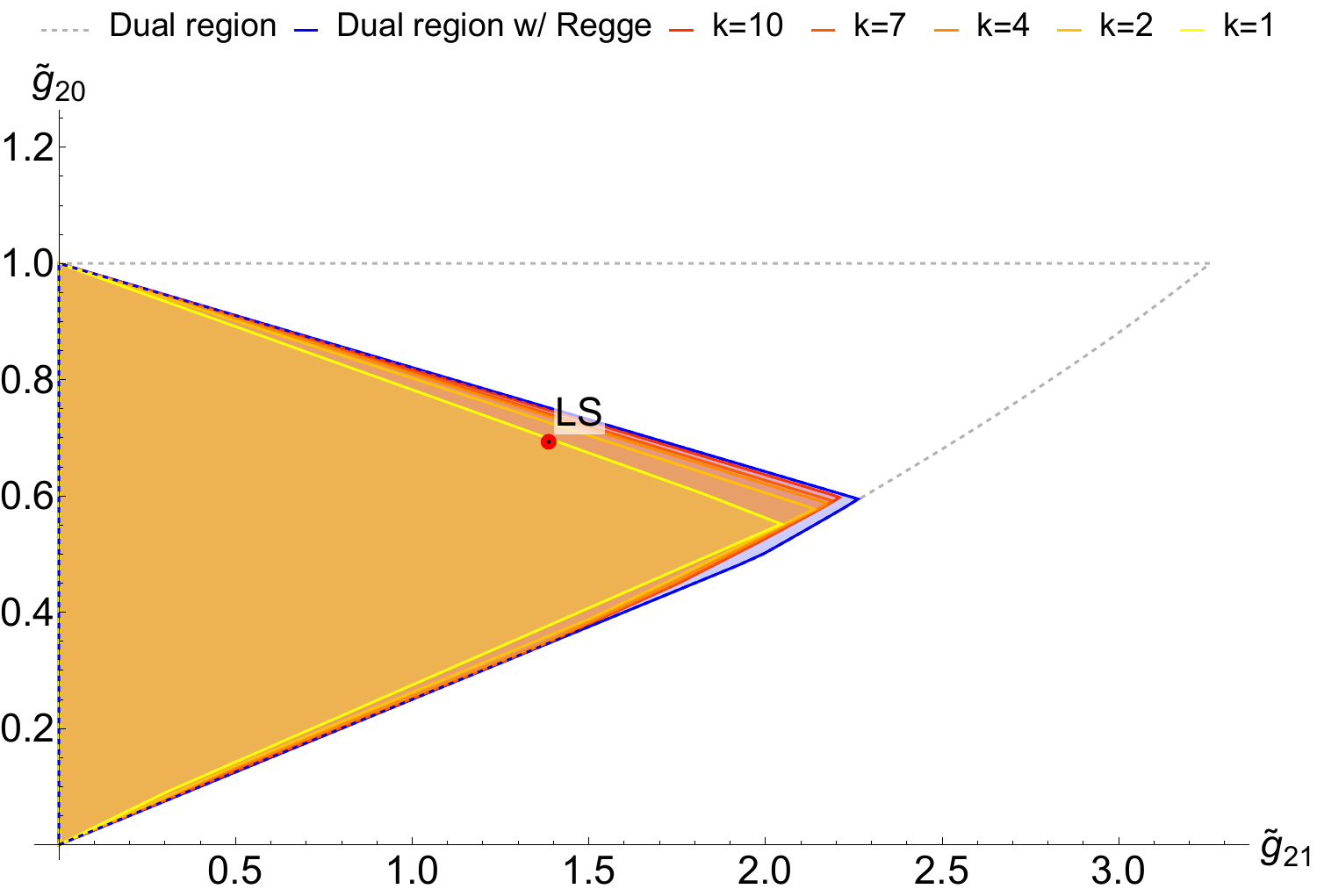}
        \caption{}
        \label{fig:LS ideal}
    \end{subfigure}
    \caption{Region spanned by GLS amplitudes  varying the upper bound on $k_\text{max}$. Left: region obtained with ansatz \eqref{Ansatz}, without corrections from pure scalar exchanges. Right: region obtained with the ansatz \eqref{Ansatz-scalars}, including pure scalar exchanges. The two regions seem to agree for large values of $k_\text{max}$.}
    \label{fig:LS comparison}
\end{figure}

By varying the parameters $c_{n,k}$ in our ansatz -- while imposing the positivity of partial waves -- we can chart the allowed values of $\tilde{g}_2$ and $\tilde{g}_2'$ defined in~\eqref{eq:g2g2p}. The outcome is shown in Figure~\ref{fig:LS no spin zero}. These ratios have also been extensively analyzed through positivity bounds \cite{Albert:2022oes,Fernandez:2022kzi,McPeak:2023wmq}, which show that the parameters must be inside the dashed region depicted in the figure. As this region was derived under the same assumptions adopted here, it is reassuring to see that the domain covered by our amplitudes falls entirely within it. One important difference, however, is that our ansatz produces only local amplitudes, with a finite number of spins at any given mass. In fact, the maximal spin is a linear function of the resonance mass, corresponding to the leading Regge trajectory. Repeating the positivity bounds with the assumption that for a given spin $J$ the mass of the resonance can only be larger than $(J-2)(m_{f_2}^2-1)+m_{f_2}^2$, produces a smaller allowed region, which shrinks as $m_{f_2}$ increases. This new region is also shown in Figure~\ref{fig:LS ideal}  as the dark blue region, labelled as Dual region with Regge (assumption). In particular the right upper corner is cut: we recall that this point was associated to the non-local amplitude $M\sim 1/(s-M^2)(u-M^2)$, which is now ruled out. This dark blue region is the allowed parameter space we should compare to.

Let us now consider the primal region filled by the family of amplitudes \eqref{Ansatz}. For $k_\text{max}=1$, our ansatz reduces to the LS amplitude, which we know lies inside the allowed region. As we increase  $k_\text{max}$, the primal region gets larger: this is expected since the higher $k_\text{max}$, the more general the ansatz becomes.\footnote{We stress that for any finite $k_\text{max}$, the ansatz~\eqref{Ansatz} produces a legitimate tree level amplitude.} In addition, the primal region seems to converge to the dual region as $k_\text{max}\rightarrow{\infty}$.\footnote{There seems to be a small sliver close to the rightmost extreme that is not covered by the primal approach. It should be noted however that convergence in the dual approach is very slow in this region, as noted in \cite{Fernandez:2022kzi}.}

For comparison we compute the primal region using the modified ansatz \eqref{Ansatz-scalars}. As shown in Figure~\ref{fig:LS ideal}, in the large $k_\text{max}$ limit the two approaches agree.

\begin{figure}[h]
    \centering
    \includegraphics[width=0.7\linewidth]{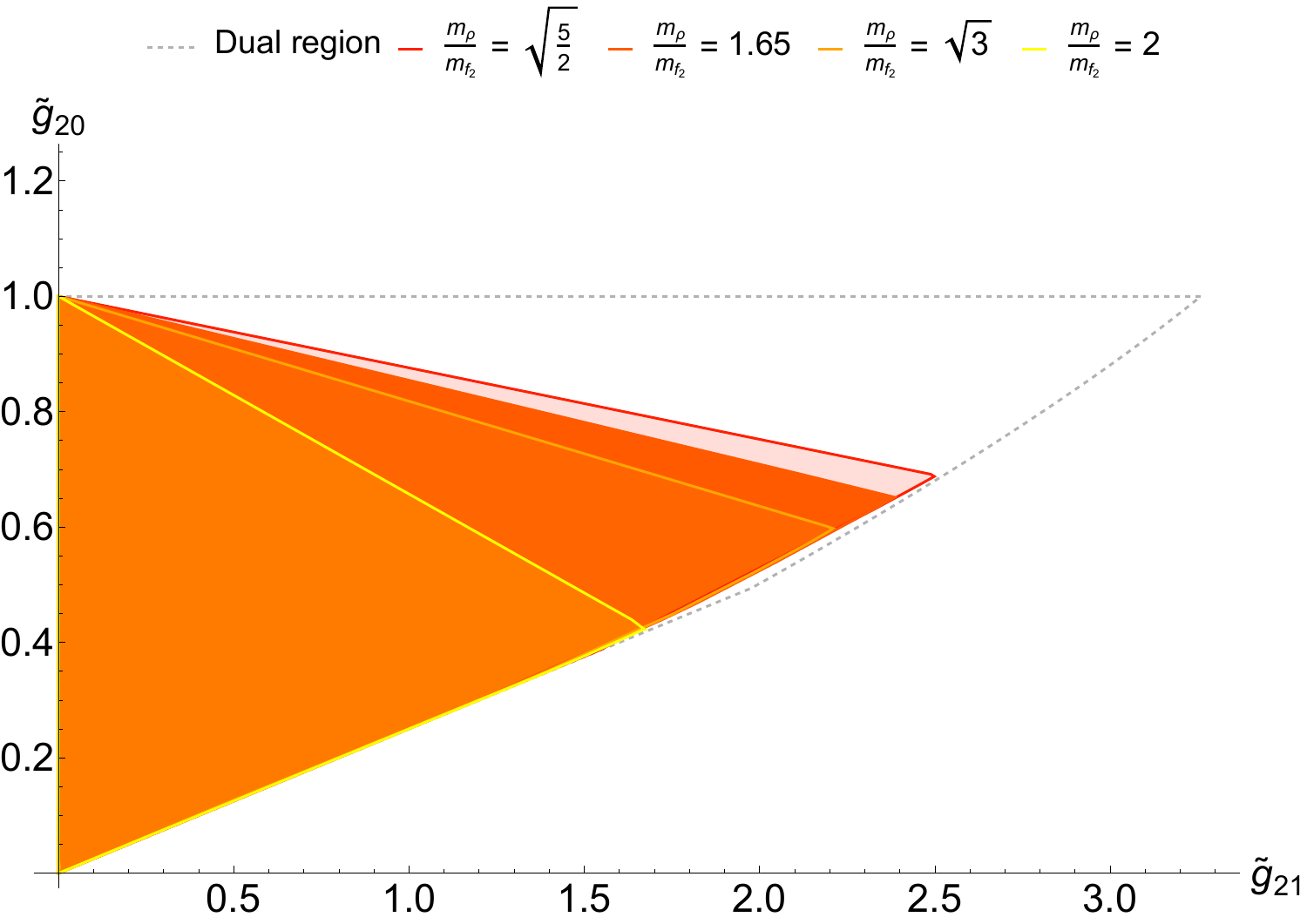}
    \caption{Region spanned by the GLS amplitude with increasing masses of the first spin-2 particle (from red to yellow). Since this mass controls the slope of the leading Regge trajectory, a larger mass corresponds to a smaller slope.}
    \label{fig:LS physical masses}
\end{figure}

Finally, we inspect how the allowed region changes as we modify the slope of the leading Regge trajectory. As expected, as the mass of the first spin-2 particle $m_{f_2}$ increases, the primal region shrinks and we checked that in each case the primal region asymptotically seem to agree with the dual allowed region obtained with  positivity bounds. This behavior is shown in Figure~\ref{fig:LS physical masses}, for values of the spin-2 resonance $m_{f_2}=\sqrt{\frac{5}{2}} \, (\text{minimal allowed value}),\, 1.65 \,(\text{physical}),\, \sqrt{3} \, (\text{standard LS}) ,2 $, in units of the leading spin-1 mass $m_{\rho}$. As reviewed in Appendix~\ref{Analytic-Study of Unitarity}, unitarity forces a  bound on the Regge slope, which translates on $m_{f_2} \geq \sqrt{5/2}$. Hence, the present ansatz does not allow to cover the whole dual region, but it is general enough to accomodate values of $m_{f_2}$ close to the experimental value.

\begin{figure}[H]
    \centering
    \includegraphics[width=0.7\linewidth]{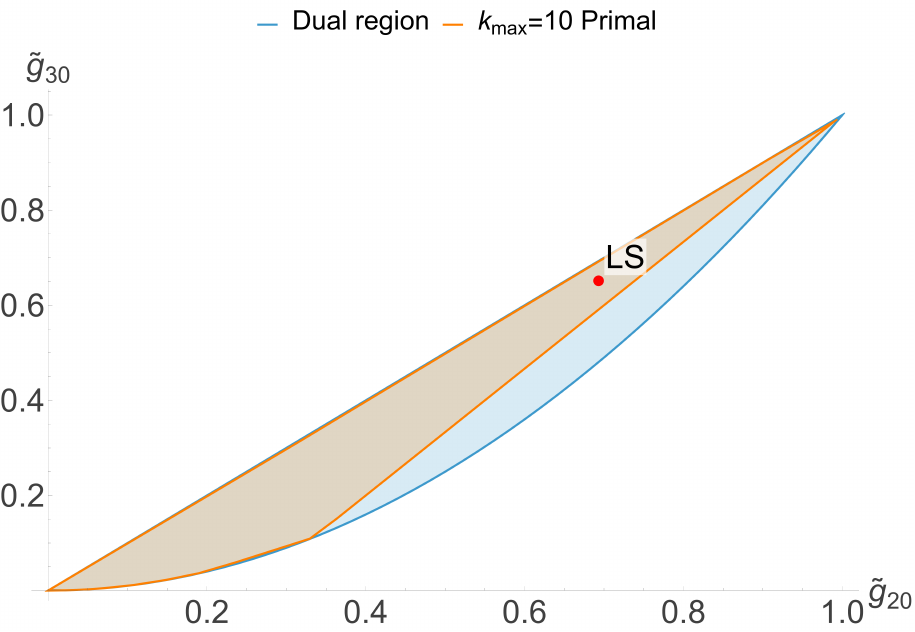}
    \caption{Region spanned by the GLS amplitude (orange) in the $(\tilde{g}_{20}, \, \tilde{g}_{3,0})$ plane. The blue region is the region allowed by dual boostrap bounds with no assumptions computed in \cite{Albert:2022oes}.}
    \label{fig:g20g30}
\end{figure}

In Figure~\ref{fig:g20g30} we considered a different combination of low energy coefficients  $(\tilde{g}_{2,1},\tilde{g}_{2,0})$. Again, the dual region with no extra assumption is larger than the parameter space covered by our ansatz, due to the existence of a leading Regge trajectory. 

On the boundary of the allowed region, the extremal solution returns the values of the coefficients $c_{n,k}$. By plugging these values in the ansatz \eqref{Ansatz} or \eqref{Ansatz-scalars}, we can also study the spectrum corresponding to these solutions. The corresponding Chew-Frautschi plots are shown in appendix \ref{Spectra of the Extremal Solutions}.

\subsection{\texorpdfstring{$\pi\pi f_2$ Coupling}
                         {pi pi f2 Coupling}}\label{pipi f2 results}
The partial-wave decomposition allows us to express our amplitude as a sum over exchanged mesons:
\begin{equation}
    M(s,t) = \sum_X g^2_{\pi\pi X} \Bigg( \dfrac{m_X^2 P_{J_X}\Big(1+\dfrac{2t}{m_X^2}\Big)}{m^2_X-s}   +(s \longrightarrow t)\Bigg)+\text{analytic}\,
\end{equation}
where we define the normalised couplings as:
\begin{equation}
    \tilde{g}_X :=\dfrac{g_{\pi\pi X}}{g_{1,0} m^2_\rho}
\end{equation}
Following \cite{Albert:2023seb}, one would like to study the three-point coupling of pions to the $f_2$ meson as a function of $1/\tilde{M}^2$, (where $\tilde{M}$ denotes the mass of the first resonance after the $f_2$ meson in units of $m_\rho$). In contrast to \cite{Albert:2023seb}, we cannot fix the $m_{f_2}$ mass at its physical value while varying $\tilde{M}$, since they are related by the leading Regge trajectory. Nevertheless, we can inspect how this coupling varies within our ansatz as a function of $m_{f_2}$ and what is the spectrum of the extremal solutions.

\begin{figure}[t]
    \centering
    \includegraphics[width=0.7\linewidth]{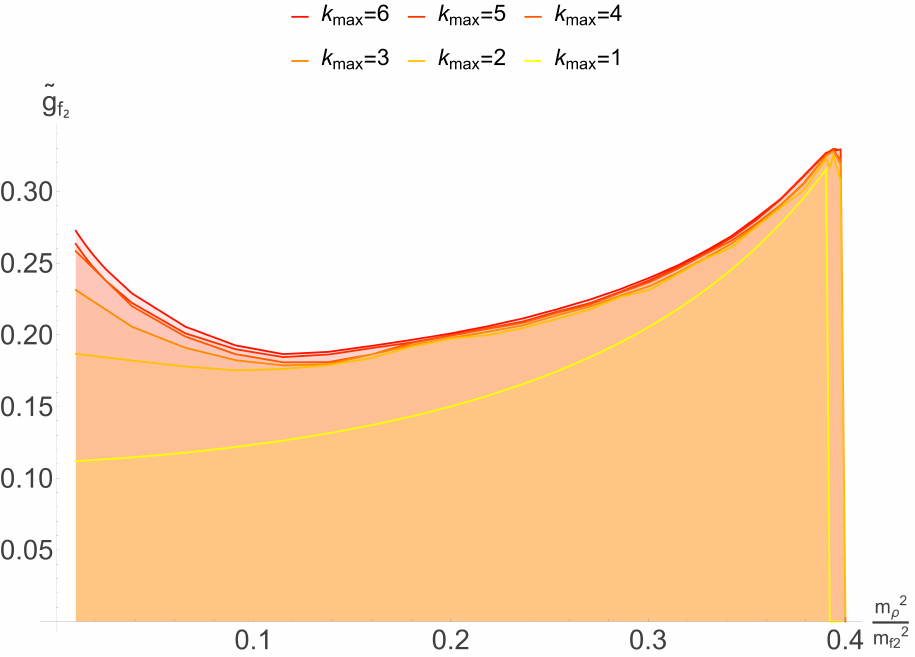}
    \caption{Maximal value of the on-shell coupling $\tilde{g}_{f_2}$ that can be obtained with the GLS amplitude with increasing values of $k_{\max}$. For $k_{\max} \geq 7$ convergence is reached for $\frac{m_\rho^2}{m^2_{f_2}} \gtrsim 0.15$.}
    \label{fig:gpipif2}
\end{figure}

As shown in figure \ref{fig:gpipif2}, the coupling converges only for higher values of $\dfrac{m_{\rho}^2}{m^2_{f_2}}$. As $m_{f_2}^2 \to \infty$,  convergence is not clear. For $k \geq 7$, the coupling appears to diverge near zero when only a finite number of constraints are imposed. Nevertheless, as the number of constraints increases, the upper bound continues to decrease. This trend persists until computational memory limitations are encountered, preventing further progress with the bootstrap algorithm.\\
In \cite{Albert:2023seb} it was found that for large values of $\tilde{M}$, the coupling $\tilde{g}_{f_2}$ must vanish. Here, however, 
we obtain a finite value, since we cannot increase the value of the mass of the first resonance after the $f_2$, while keeping $m_{f_2}$ fixed. 
For the physical value of the masses $\frac{m^2_{\rho}}{m_{f_2}^2} \simeq 0.367$, we have obtained an optimal value of:
\begin{equation}
    \tilde{g}_{f_2} \simeq 0.296
\end{equation}
This is not compatible with the experimental result for real-world QCD: $0.335^{+0.013}_{-0.007}$. However, this result is also related to the fact that fixing the masses of the first two mesons, we are fixing also the mass of the whole spectrum. For real-world QCD the masses follow an approximately linear Regge trajectory, and our approach is not able to accomodate all the features of the theory, at least for $N_c=3$.

\section{Fixed angle scattering}\label{Fixed angle scattering}

\subsection{Fixed-Angle High-Energy Limit in QCD}
While, for $s \gg -t \gg \Lambda_{\mathrm{QCD}}$, the experimental data behave in accordance with Regge theory, for $s \sim -t \gg \Lambda_{\mathrm{QCD}}$, the amplitude must exhibit a power-like suppression with a constant power that does not carry any $t$-dependence. Historically, this has been imposed by requiring the Regge trajectory to approach the correct negative constant asymptotically (see, for example, \cite{Collins1984}). Here we show that, at least in the large-$N_C$ framework, the Regge limit and the fixed-angle high-energy limit do not automatically commute.\\ 
In the high-energy limit, with fixed $s/t$ ratio, the scattering amplitude has a universal behavior. If both $s,t >0$ we expect the amplitude to explode exponentially, following the behavior of the Veneziano amplitude \cite{Caron-Huot:2016icg}:
\begin{equation}
 M(s,t) \sim e^{\alpha \left(s\log\left(\frac{s+t}{s}\right)+t\log\left(\frac{s+t}{t}\right)\right)}  
 \label{FA limit unphysical}
\end{equation}
While in the physical region ($s>0$, $t<0$), we expect the polynomial suppression given by the Brodsky–Farrar counting rules \cite{Brodsky.31.1153}:
\begin{equation}
 M(s,t) \sim s^{2-\frac{n_q}{2}}F(\theta)  
  \label{FA limit physical}
\end{equation}
where $n_q$ is the number of valence quarks involved in the process and $F(\theta)$ is some angle-dependent function.\\

As discussed in \cite{Polchinski:2001tt}, gauge–string duality arguments suggest that an exponential–power-like transition may occur at negative $t$. A similar behavior was previously described in \cite{Collins1984}, where the large negative-$t$ region of the Regge limit was made compatible with the fixed-angle high-energy region through a Reggeisation of internal gluon exchanges.
Here we have followed the alternative method outlined in \cite{Veneziano:2017cks}, analytically studying both the fixed-angle high-energy behavior of the amplitude and its Regge behavior. We have found that the Brodsky–Farrar counting rules can be imposed through the same mechanism outlined in Veneziano’s paper. The exchange of dominance between different Regge trajectories, which is the key to restoring the polynomial behavior, introduces an apparent non-linear leading trajectory:
\begin{equation}
    \lim_{t \to -\infty} \alpha(t)=\text{const}
\end{equation}
In our model, the leading behavior is given by an infinite number of Regge trajectories containing the $\rho$ meson and therefore contributing as $M(s,t) \sim s^1$. By construction the Regge limit and the fixed-angle high energy limit do not commute in our model and, when the $s/t$ ratio is held fixed, our model will exhibit the expected power-like behavior.

\subsection{Fixed-Angle High-Energy Limit in the GLS model}\label{Fixed-Angle High-Energy Limit}
Each amplitude obeying the properties required for large-$N$ QCD, in the high-energy limit $s,t \to +\infty$, with their ratio kept fixed, should satisfy the following behavior \cite{Caron-Huot:2016icg}:
\begin{equation}\begin{split}
    M(s,t) \sim  e^{ (\alpha'((s+t)\log(s+t)-s\log(s)-t\log(t))) }\equiv  \exp (s f(s/t))
\end{split}\end{equation}
for some constant $\alpha'$.\\
This behavior does not match the Brodsky–Farrar rules \cite{Brodsky.31.1153,Veneziano:2017cks}, required in the physical limit. As discussed in \cite{Veneziano:2017cks}, some form of Stokes phenomenon can restore the behavior expected from QCD in the physical limit ($s,-t \to +\infty$). The explicit computation, following the procedures outlined in \cite{Veneziano:2017cks}, has been done in Appendix~\ref{High-Energy Fixed-Angle Limit appendix}, here we show the main results.\\
The dominant contribution to the amplitude in the unphysical region is given once again by the Lovelace–Shapiro amplitude, for any general parametrization of the coefficients $c_{n,k}$.\\
If the sum is bounded, the leading behavior will be that of the amplitude with $k=k_{\max}$.\\
It is more interesting to evaluate the amplitude when $k_{\max}$ is taken to infinity. To estimate the sum, we must introduce an asymptotic behavior for the free coefficients of our theory, and, as before, we assume a polynomial suppression:
\begin{equation}
     c_{0,k} \sim \dfrac{1}{k^{\gamma}}
    \label{FA limit- coefficients behavior}
\end{equation}
With this assumption we recover the Brodsky-Farrar behavior:
\begin{equation}
    M(s,t) \sim F(\theta)\, s^{1-\gamma}
\end{equation}
More precisely, the Brodsky-Farrar rules \eqref{FA limit physical}, for ordinary QCD (with $n_q=8$), correspond to the choice $\gamma=3$.

\subsection{Imposing the Fixed-Angle High-Energy behavior}
\label{Imposing FA constraint}

As discussed in Section~\ref{Fixed-Angle High-Energy Limit}, the Brodsky--Farrar behavior
\begin{equation}
    M(s,t)\sim F(\theta)\,s^{1-\gamma}
\end{equation}
with $\gamma=3$ for ordinary QCD, is not a property of any individual amplitude
$A_{n,k}$ at fixed $k$. Rather, it arises from an infinite resummation over the large-$k$
tail of the ansatz. Indeed, at fixed angle, with
\begin{equation}
    t=\frac{s}{2}(z-1),\qquad -1<z<1,
\end{equation}
each fixed-$k$ contribution is exponentially suppressed,
\begin{equation}
    A_{n,k}(s,z)\sim 
    \exp\left[-\frac{s}{k}F(z)\right],
    \qquad F(z)>0.
\end{equation}
Thus, for any finite truncation $k\leq k_{\max}$, the asymptotic fixed-angle behavior is
dominated by the largest available value of $k$ and remains exponentially suppressed. The
power-law behavior can only emerge if the sum includes terms with
\begin{equation}
    k\sim s
\end{equation}
as $s\rightarrow\infty$. In this regime the large-$k$ expansion gives
\begin{equation}
    A_{n,k}(s,z)\sim -\frac{2k(1+z)}{(1-z)s},
\end{equation}
and therefore, if
\begin{equation}
    c_{n,k}\sim \frac{c_n}{k^\gamma},
\end{equation}
the tail contributes
\begin{equation}
    M_{\rm tail}(s,z)
    \sim
    \frac{1}{s}\sum_{k\gtrsim s} k^{1-\gamma} \simeq \frac{\zeta (\gamma -1,s)}{s}
    \sim
    s^{1-\gamma}.
\end{equation}
This is the mechanism by which the Brodsky--Farrar scaling is recovered in the GLS ansatz.

It is useful to translate this statement into the language of the positivity constraints. The
poles of $A_{n,k}$ occur when
\begin{equation}
    \alpha_{n,k}(s)=q+1,
    \qquad q=0,1,2,\ldots,
\end{equation}
or equivalently
\begin{equation}
    \alpha' s=n+1-\alpha_0+kq.
\end{equation}
In terms of the common mass-level variable
\begin{equation}
    N=\alpha' s+\alpha_0-1,
\end{equation}
this gives
\begin{equation}
    N=n+kq.
\end{equation}
The residue at this pole is a polynomial of degree of order $q+1$ in the angular variable, and
therefore contributes only to spins $J\lesssim q+1$. Consequently, the large-$k$ region relevant
for fixed-angle scattering should not be thought of primarily as a high-spin region. Rather, for
$k\sim N$, it corresponds to arbitrarily high mass levels but only finite or moderate spin. Thus,
the obstruction in the numerical implementation is not that increasing $k_{\max}$ necessarily
forces us to include much higher spins; it is that the Brodsky--Farrar tail lives at
\begin{equation}
    N\sim k\sim s\rightarrow\infty.
\end{equation}
A finite truncation of the positivity constraints,
\begin{equation}
    J\leq N\leq N_{\max},
\end{equation}
cannot certify the unitarity of this asymptotic tail.

We nevertheless attempted to impose the fixed-angle scaling directly at the level of the finite
numerical ansatz. Instead of treating all coefficients $c_{n,k}$ independently, we restricted them
to the form
\begin{equation}
    c_{n,k}
    =
    \sum_{\ell\in\mathcal L}
    \frac{b_{n,\ell}}{(k+\ell)^\gamma}.
    \label{eq:fixed-angle-trial-ansatz}
\end{equation}
For any finite set $\mathcal L$, this is a sufficient, though not necessary, way of producing
$c_{n,k}\sim k^{-\gamma}$ if the expression is continued to arbitrarily large $k$. In our the numerical investigations we used $ \mathcal L=\{0,1,2,3,4,5,10\}$, and truncated the sum at finite $k_{\max}$.

This procedure is useful as a diagnostic, but it should not be interpreted as a controlled
implementation of the Brodsky--Farrar constraint. There are three related reasons.

First, the fixed-angle power law is an asymptotic statement about the region $k\sim s$ as
$s\rightarrow\infty$. At any finite $k_{\max}$, the terms responsible for the true fixed-angle
limit are absent once $s\gg k_{\max}$. The finite computation therefore probes only a
pre-asymptotic interpolation, not the actual asymptotic tail.

Second, the parametrisation~\eqref{eq:fixed-angle-trial-ansatz} imposes correlations between
low, intermediate, and large values of $k$. However, the Brodsky--Farrar condition constrains
only the large-$k$ tail. There is no physical reason for the coefficients at small $k$, which control
much of the low-energy data, to be tied rigidly to the coefficients at asymptotically large $k$.
Therefore, any reduction of the allowed region obtained from
\eqref{eq:fixed-angle-trial-ansatz} can partly reflect the reduced flexibility of the finite-dimensional
trial family, rather than a genuine consequence of fixed-angle scaling.

Third, even if the coefficients are forced to have the desired large-$k$ behavior within the chosen
parametrisation, the positivity constraints are still imposed only up to finite $N_{\max}$. Since
the fixed-angle tail probes $N\sim k\rightarrow\infty$, the numerical program does not prove that
the imposed tail is compatible with unitarity at all mass levels.
%

Numerical investigations shows that the allowed region in the
$(\widetilde g_{2,1},\widetilde g_{2,0})$ plane is only mildly reduced. This is the behavior one should expect on general grounds. The fixed-angle condition is a constraint on the asymptotic
UV completion of the amplitude, while the low-energy EFT coefficients are determined by the
Taylor expansion around $s=t=0$. If the normalization of the fixed-angle tail is not fixed, one
can add a tail of the schematic form
\begin{equation}
    \delta c_{n,k}
    =
    \varepsilon\, d_{n,k}\,\Theta(k-K),
    \qquad
    d_{n,k}\sim k^{-\gamma},
\end{equation}
with $K$ arbitrarily large and $\varepsilon$ arbitrarily small. Such a tail can determine the strict
asymptotic fixed-angle behavior, because all finite-$k$ contributions are exponentially suppressed
at sufficiently large $s$, while its effect on any fixed number of low-energy coefficients can be
made arbitrarily small. Hence the Brodsky--Farrar condition certainly restricts the space of full
amplitudes, but it need not produce a sizeable restriction after projecting onto a small set of
low-energy EFT data. A substantial reduction of the low-energy allowed region would require
additional information, such as a fixed nonzero normalization of $F(\theta)$ or a finite onset scale
above which the power-law behavior must already be visible.\footnote{A more appropriate treatment of the fixed-angle constraint should therefore separate the IR part
of the ansatz from the UV tail. One possible strategy is to write
\begin{equation}
    c_{n,k}
    =
    c^{\rm IR}_{n,k}\,\Theta(K-k)
    +
    c^{\rm UV}_{n,k}\,\Theta(k-K),
    \qquad
    c^{\rm UV}_{n,k}
    =
    k^{-\gamma}
    \sum_r b_r\,\phi_r\!\left(\frac{n}{k}\right),
\end{equation}
where the first term is optimized freely, while the second term is designed to reproduce the
large-$k$ fixed-angle scaling. The positivity constraints could then be imposed with the present
algorithm up to finite $N_{\max}$ and supplemented by a cutting-plane search for violations at
larger $N$, or by an analytic large-$N$ study of the UV tail.}

Another possibility is to study observables that are more directly sensitive to the fixed-angle
regime than the first few EFT coefficients. Examples include finite-energy fixed-angle quantities,
such as
\begin{equation}\label{eq:new-observable1}
    \mathcal B_\gamma(S,z)
    =
    S^{\gamma-1}
    M\!\left(S,\frac{S}{2}(z-1)\right),
\end{equation}
evaluated away from poles or after a mild smearing in $S$, and ratios such as
\begin{equation}\label{eq:new-observable2}
    \mathcal R_\gamma(S_1,S_2;z)
    =
    \frac{M\!\left(S_2,\frac{S_2}{2}(z-1)\right)}
         {M\!\left(S_1,\frac{S_1}{2}(z-1)\right)}
    \left(\frac{S_2}{S_1}\right)^{\gamma-1}.
\end{equation}
For a true power law these quantities approach finite limits, whereas for a finite-$k$ truncation
they retain exponential dependence on $S/k_{\max}$. Since these observables are linear or ratios
of linear functionals of the coefficients, they can be studied with essentially the same primal
bootstrap machinery.

Finally, one can probe the same physics spectrally. The fixed-angle tail corresponds to a
characteristic distribution of pion couplings at large mass level and low or moderate spin. Thus,
instead of projecting immediately onto low-energy EFT coefficients, one may optimize moments
of the positive partial-wave residues, for instance
\begin{equation}\label{eq:new-observable3}
    \mathcal W_{p,J_0}(N_0)
    =
    \sum_{N>N_0}\sum_{J\leq J_0}
    N^p\,\rho_{N,J},
\end{equation}
or study the large-$N$ falloff of the residues along the low-spin sector. In spectral language, the Brodsky--Farrar tail corresponds to a power-law distribution of pion
couplings to high-mass, low-spin resonances. One expects that the physical squared pion couplings scale as
\[
    g^2_{\pi\pi;M^2,J}\sim M^{-2\gamma},
\]
up to an $O(1)$ spin-dependent coefficient.

The alternative observables \eqref{eq:new-observable1}, \ref{eq:new-observable2} and \eqref{eq:new-observable3} are
better adapted to distinguishing amplitudes with a genuine Brodsky--Farrar tail from amplitudes
whose fixed-angle behavior remains exponentially suppressed at any finite $k_{\max}$. We leave the investigation of these quantitites for future works.

\section{Conclusions}
\label{Conclusions}

In this work, we have developed and examined in detail a class of meromorphic scattering amplitudes derived from the classical Lovelace--Shapiro model. The explicit solutions obtained in this construction simultaneously satisfy the fundamental $S$-matrix requirements of analyticity, crossing symmetry, and tree-level unitarity, the latter being imposed through positivity of the partial-wave residues. Our construction was designed to capture the key properties expected from large-$N_c$ QCD, in particular the characteristic Regge behavior at high energy and fixed momentum transfer. Furthermore, a specific asymptotic tuning of the free parameters of our model is compatible with a polynomial suppression in the high-energy fixed-angle limit, as expected from the Brodsky--Farrar counting rules.

We subsequently enforced unitarity in our model using standard $S$-matrix bootstrap techniques, deriving bounds on the quadratic coefficients of the low-energy expansion of the amplitude and on the coupling of the $\pi\pi f_2$ interaction. The resulting primal regions are compatible with the corresponding dual positivity bounds, and in the regimes where the numerics converge they appear to fill the expected allowed regions. This provides evidence that the generalized Lovelace--Shapiro basis is sufficiently flexible to capture a large class of tree-level large-$N_c$ amplitudes obeying the assumptions imposed in this work.

Our spectral analysis, see Figure~\ref{fig:Spectrum Evolution near the kink}, suggests that a qualitative transition occurs in the couplings of the mesons to the pions near the right kink. Approaching the kink along the boundary, the dominant spectral weight reorganizes: below the kink the pion couplings are distributed over a broad set of resonances, while near and above the kink they become increasingly concentrated in a pattern closer to that of the Lovelace--Shapiro amplitude. This observation suggests that the kink is not merely a numerical feature of the projected EFT space, but may reflect a change in the organization of the underlying resonance spectrum.

Finally, the numerical study of the high-energy fixed-angle behavior has shown that the optimal bounds are often saturated by the first few terms of the ansatz, leaving to the higher-order terms the freedom to accommodate some high-energy features of the theory, only marginally affecting the IR effective amplitude. This mild effect should not be surprising. The Brodsky--Farrar behavior is an asymptotic statement about the large-$k$ tail of the coefficients, with the dominant terms appearing at $k\sim s$ in the fixed-angle limit. 

This suggests several directions for future work. A first possibility is to study the fixed-angle limit using observables that are more directly sensitive to it than the first few Taylor coefficients around $s=t=0$, as suggested in Section~\ref{Imposing FA constraint}.

It would also be interesting to formulate dispersion relations adapted to fixed scattering angle, or to use angle-smeared dispersive functionals that interpolate between standard fixed-$t$ dispersion relations and the fixed-angle regime. Such a formulation could make the Brodsky--Farrar condition a direct constraint on the amplitude, rather than an indirect condition on the asymptotic large-$k$ behavior of the coefficients. 

A second direction is to enlarge the space of building blocks. The present basis is constructed from Lovelace--Shapiro--like amplitudes with linear Regge trajectories. Other dual amplitudes, such as the Coon amplitude and related accumulation-point amplitudes, provide examples with logarithmic trajectories, accumulation points in the spectrum, and different fixed-angle behavior~\cite{FigueroaTourkine2022,MaldacenaRemmen2022}. Studying analogous primal bootstrap problems in such bases could clarify which aspects of our results are consequences of meromorphy, crossing and unitarity, and which depend on the assumption of approximately linear Regge trajectories. More generally, it would be valuable to construct new meromorphic bases that allow for non-linear Regge behavior while preserving crossing symmetry and a tractable partial-wave expansion. 

Finally, one could combine the present primal approach with neural-network parametrizations of the missing data. Instead of choosing a finite set of coefficients $c_{n,k}$, or a rigid analytic ansatz for their large-$k$ tail, one could parametrize the coefficients, the spectral density, or the Regge trajectories themselves by neural networks and train them with a loss function imposing crossing, partial-wave positivity, Regge behavior, and fixed-angle scaling. Recent neural approaches to the $S$-matrix bootstrap \cite{GumusLeflotTourkineZhiboedov2024,GumusLeflotTourkineZhiboedov2026} and conformal bootstrap \cite{Ghosh:2026xnp,Ghosh:2026jbw,Benjamin:2026lbj} suggest that such methods can provide flexible parametrizations of amplitudes while still allowing one to impose analyticity and unitarity constraints numerically. In this way, one could impose all sort of condition on the amplitude with a suitable choice of the cost function.

These extensions would help separate three logically distinct questions: how much of the low-energy pion amplitude is fixed by analyticity, crossing and unitarity alone; how much is fixed by imposing Regge behavior; and which observables are genuinely sensitive to the high-energy fixed-angle tail.

\section*{Acknowledgments}

We thank J. Albert, J. Henriksson, L. Rastelli and A. Guerrieri for valuable discussions.
AV is partially supported by the Italian MUR under the project 20223ANFHR (PRIN2022).
SB is supported by the Royal Society through University Research Fellowship URF/R1/241371.

\appendix

\section{Analytic-Study of Unitarity}\label{Analytic-Study of Unitarity}
\subsection{Analytic-Study of the High-Energy Poles}
In some models (for instance, the amplitude considered in \cite{Haring:2023zwu}) it is not possible to truncate the two sums in \eqref{Ansatz} and impose an upper bound on the value of $k_{\max}$. We will follow the procedure outlined in Appendix D of \cite{Haring:2023zwu} for our ansatz and compare the results.\\
Here we consider a fixed slope $\alpha'=1$ in \eqref{Ansatz}; the proof for the general case is analogous, with the substitution:
\begin{equation}
    s,t \to \alpha' s,\alpha' t
\end{equation}
The issue with the amplitude considered in \cite{Haring:2022cyf} is that the coefficients of the spin $n-k$ and $n-k-1$ partial waves for a fixed $\{n,k\}$ amplitude have opposite signs. We now show that this problem does not appear in the model that we have proposed in \eqref{Ansatz}.\\
The highest spin that can be exchanged on-shell at a fixed energy is that contained in the leading trajectory. Therefore, in this case only one term contributes to the partial wave. We check that, in a general $\{n,k\}$ amplitude, the sign problem for the two highest-spin terms does not arise. Let us consider a generic LS amplitude with $\alpha(s)=\alpha_0+\alpha' s$. The residue at $\alpha(s)=m$ takes the form:
\begin{equation}\begin{split}
    \mathrm{Res}[A_{LS}(s,t)\,|\,\alpha(s)=m] 
    &= \dfrac{(-1)^m}{(m-1)!}\dfrac{\Gamma(1-\alpha(t))}{\Gamma(1-\alpha(t)-m)}\\
    &= \dfrac{(-1)}{(m-1)!}\Big[\alpha(t)^{m}+\dfrac{(m-1)m}{2}\alpha(t)^{m}+...\Big]\\
    &= \dfrac{(-1)}{(m-1)!}\Bigg[ \Big(\dfrac{m-\alpha_0}{2}\Big)^{m}x^{m} \\
    &\quad+\dfrac{m}{2^m} \left[(m+1)(m+\alpha_0)^{m-1}-(m-3\alpha_0)(m-\alpha_0)^{m-1}\right] x^{m-1}+...\Bigg]
\end{split}\end{equation}
The coefficient of the term in $x^{m-1}$ is always positive for intercepts $0 < \alpha_0 < 1$, and hence the coefficients in the partial-wave expansion for the two highest spins have the same sign.\\
The possibility of considering a finite $k_{\max}$ will be useful in our numerical study. In fact, in developing our bootstrap algorithm we will always consider a truncated ansatz.

\subsection{Analytic-Study of the Low-Energy Poles}
If there is no issue in considering a single amplitude at high energies, the poles at low energy exhibit a non-trivial behavior.\
The question we address is whether the individual amplitudes entering the GLS sum are independently unitary.\\
Let us first consider a single Lovelace--Shapiro amplitude with Regge trajectory
\begin{equation}
\alpha(s)=\alpha_0+\alpha' s .
\label{Regge trajectories}
\end{equation}
The first pole gives two constraints, one for the spin-0 and one for the spin-1:
\begin{equation}
\dfrac{-1+3\alpha_0}{2\alpha'} \geq0 \qquad
\dfrac{1-\alpha_0}{\alpha'}\geq0 \, .
\label{PW1}
\end{equation}
At the second pole we also find a spin-2 contribution:
\begin{equation}\begin{split}
\dfrac{(-1 + 2 \alpha_0) (-2 + 7 \alpha_0)}{6\alpha'} \geq0,\qquad
\dfrac{-2+7\alpha_0-3 \alpha_0^2}{2\alpha'} \geq0 ,\qquad
\dfrac{(\alpha_0-2)^2}{6\alpha'}\geq0\, .
\label{PW2}
\end{split}\end{equation}
Requiring the single amplitude to be unitary by itself gives
\begin{equation}
\dfrac{1}{2} \leq \alpha_0 \leq 1 \, .
\end{equation}
In terms of the lightest spin-1 and spin-2 particles,
\begin{equation}
\alpha_0=1-\dfrac{m_1^2}{m_2^2-m_1^2} ,\qquad
\alpha'=\dfrac{1}{m_2^2-m_1^2} \, .
\end{equation}
The lower end $\alpha_0=\tfrac12$ corresponds to the Adler-zero value of the ordinary Lovelace--Shapiro amplitude \cite{PhysRev.137.B1022}. If the intercept falls below this value, the first obstruction is a negative scalar residue. This obstruction is not fundamental in the enlarged ansatz, since adding spin-0 exchanges does not violate crossing, meromorphy or Regge boundedness \cite{Albert:2022oes,Fernandez:2022kzi}. In this case the scalar constraints in \eqref{PW1} and \eqref{PW2} can be repaired independently.

After removing the scalar obstruction, the first non-scalar condition for a single trajectory comes from the spin-1 residue at the second pole,
\begin{equation}
\dfrac{-2+7\alpha_0-3\alpha_0^2}{2\alpha'}
=
\dfrac{(2-\alpha_0)(3\alpha_0-1)}{2\alpha'}
\geq0 \, .
\end{equation}
For $0<\alpha_0<1$ this gives $\alpha_0\geq \tfrac13$, or equivalently
\begin{equation}
m_2^2 \geq \dfrac{5}{2}m_1^2 \, .
\end{equation}
This bound, however, is still only a bound on an isolated LS trajectory supplemented by scalar exchanges. It is not a bound on the full GLS ansatz. Indeed, the GLS amplitudes are constructed so that different terms in the sum contribute to the same physical mass levels. For example, the negative spin-1 residue of the $(n,k)=(0,1)$ term at the second pole can be compensated by the first pole of the $(n,k)=(1,k)$ terms.  Therefore a negative contribution from the leading LS block for $\alpha_0<\tfrac13$ need not violate unitarity if it is compensated by the other GLS coefficients. In practice, however, we see numerically however that unitarity is never satisfied with a finite $k_\text{max}$. In this work we then alwasys restric to $\alpha_0 \geq 1/3$.

\section{Asymptotic behavior of the Amplitude} \label{Alternative Generalisation of the Amplitude and Dependence by the Coefficients Asymptotic behavior}

\subsection{Regge Limit}\label{Regge Limit appendix}
\begin{figure}[H]
    \centering
    \begin{subfigure}[b]{0.45\textwidth}
        \centering
        \includegraphics[width=\textwidth]{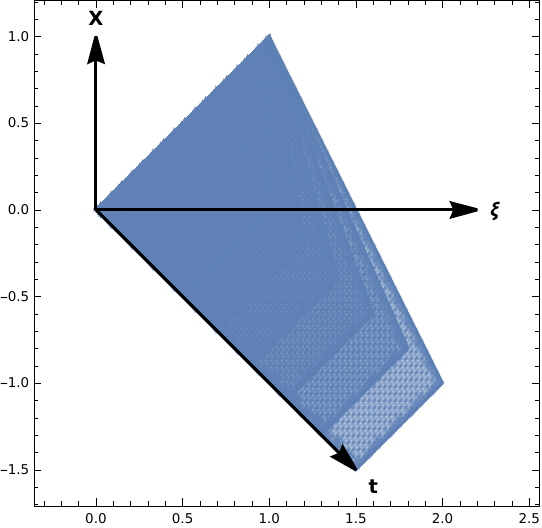}
        \caption{Positive $t$}
        \label{fig: prima regione integrazione}
    \end{subfigure}
    \hfill
    \begin{subfigure}[b]{0.45\textwidth}
        \centering
        \includegraphics[width=\textwidth]{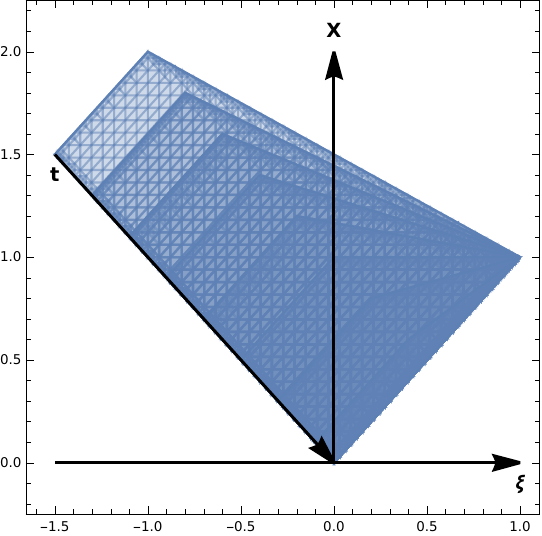}
        \caption{Negative $t$}
        \label{fig: seconda regione integrazione}
    \end{subfigure}
    \caption{Integration region for positive and negative $t$.}
    \label{fig:regione integrazione}
\end{figure}

Here we present the explicit computation that leads to the results shown in section \ref{Regge Limit}. We want to evaluate the sum in equation \eqref{Regge limit 1} in the large $s$ limit. The leading term in $n$ is always given by the lowest $n$ ( which is $n=0$ for most of the physical situations that we are interested in). We can then approximate the infinite sum in $k$ as an integral, following the procedure of \cite{Veneziano:2017cks}:
\begin{equation}
    \dfrac{M(s,t)}{g^2} \sim \sum_{k=1}^{\infty} c_{0 ,k}\, \Bigg(-\frac{\alpha'}{k}s\Bigg)^{ \frac{k-1+\alpha_0}{k}+\frac{1}{k} \, \alpha' \, t}
\end{equation}
Changing the integration variable ($k\to 1/\lambda$),
\begin{equation}
   M(s,t)\sim \int_0^1 \dfrac{d\lambda}{\lambda^2} c \Bigg(0,\frac{1}{\lambda}\Bigg) s^{1+(\alpha_0-1+\alpha' \, t)\lambda } 
\end{equation}
For $\alpha' t > 1-\alpha_0$ the main contribution to the integral is given by $\lambda \approx 1$\footnote{One might be concerned about a possible divergence near $\lambda=0$. However, an explicit computation of this contribution (which dominates in the $\alpha' t < 1-\alpha_0$ region) shows that its contribution is subleading.}. In this limit the leading behavior of the amplitude is controlled by the Regge trajectory $\alpha_L=\alpha_0+\alpha' \, t$ of the Lovelace–Shapiro model.\\
To evaluate the other regime $\alpha' t < 1-\alpha_0$, we assume a power-like behavior:
\begin{equation}
    c_{0,k} \sim \dfrac{1}{k^{\beta+1}}
\end{equation}
The leading behavior in the Regge limit is:
\begin{equation}
   M(s,t)\sim \int_0^1 \dfrac{d\lambda}{\lambda^{1-\beta}}  s^{1+(\alpha_0-1+\alpha' \, t)\lambda } 
   \label{integral regge behavior}
\end{equation}
In this region the dominant contribution to an integral of the form \eqref{integral regge behavior} comes from small $\lambda$. Therefore, we can neglect the lower bound on $k$ (and hence the upper bound on $\lambda$):
\begin{equation}\begin{split}
    M(s,t)\sim  s \int_0^{+\infty} \dfrac{d\lambda}{\lambda^{1-\beta}}  e^{(\alpha_0-1+\alpha' \, t)\log(s) \lambda } \\
    =\,s \big[(1-\alpha_0-\alpha' \, t)\log(s)\big]^{-\beta} \int_0^{+\infty} dx \, x^{\beta-1}e^{-x}
    \label{Regge negative t integral}
\end{split}\end{equation}
The integral converges for $\beta>0$, and we obtain the result shown in Section~\ref{Regge Limit}:
\begin{equation}
    M(s,t) \sim s \log^{-\beta}(s).
    \label{Regge negative t with log}
\end{equation}
We can now generalize our procedure, for a more general situation.\\
As we have seen, the amplitude obeys the principles of Regge phenomenology: the spectrum controls the behavior in the limit $s \to \infty$ at fixed $t$. However, the behavior at negative $t$ does not depend on the spectrum. A similar phenomenon will occur in the high‑energy fixed‑angle limit, both in the physical and in the unphysical regions.\\
We can parametrize the most general amplitude as an integral of LS-amplitudes, over different intercepts and slopes:
\begin{equation}
    M(s,t) \sim \int_0^{\alpha'} da_1 \int_0^{1-a_1} da_0 \, \mu(a_1,a_0)\, M_{LS}(s,t;a_1,a_0)
    \label{More general ansatz}
\end{equation}
where $\mu(a_1,a_0)$ denotes the weight of the integration and where we have fixed the mass of the $\rho$ meson to $m_\rho=1$. Usually, one would assume a discrete measure $\mu(a_1,a_0)$ selecting a precise spectrum; however, we have seen that in the non‑trivial region the dominant terms become asymptotically dense.\\
We start now to compute the Regge limit of this model:
\begin{equation}
    M(s,t) \sim \int_0^{\alpha'} da_1 \int_0^{1-a_1} da_0 \, \mu(a_1,a_0)\, s^{a_0+a_1 t}
\end{equation}
From the change of variables $k,m \to a_1,a_0$, we expect an additional factor $a_1^3$, that we show explicitly by redefining the weight:
\begin{equation}
    M(s,t) \sim \int_0^{\alpha'} da_1 \int_0^{1-a_1} da_0 \, a_1^3 \, \mu_R(a_1,a_0)\, s^{a_0+a_1 t}
\end{equation}
We now perform the change of variables:
\begin{equation}
    \begin{split}
        \xi &= a_0+a_1 t\\
        X   &= a_0-a_1 t
    \end{split}
\end{equation}
For positive $t$, the integration trapezoid lies entirely in the positive $\xi$ region, while for negative $t$ the integration contour lies in the trapezoid reflected along the $X=\xi$ axis, as shown in figure \ref{fig:regione integrazione}.
Let us compute the integral in the positive-$t$ region:
\begin{equation}
    M(s,t) \sim \int_{0}^{\max[1,1+(t-1)\alpha']}d\xi \int_{f_l(\xi)}^{f_u(\xi)} dX \, \tilde{\mu}(\xi,X)\, s^\xi
\end{equation}
where $f_l(\xi)$ and $f_u(\xi)$ are the parametrisations of the contour.\\
For negative $t$, the integral becomes:
\begin{equation}
    M(s,t) \sim \int_{\alpha' t}^{1}d\xi \int_{f_l(\xi)}^{f_u(\xi)}dX \, \tilde{\mu}(\xi,X)\, s^\xi
\end{equation}
We now show that assuming that the weight factorizes $\tilde{\mu}(\xi,X)\approx \tilde{\mu}_1(\xi)\tilde{\mu}_2(X)$, gives us the expected result in the positive $t$ region and a result analogous to \eqref{Regge negative t} for negative $t$'s.
\begin{equation}
    M(s,t) \sim \sim \int_{0}^{\max[1,1+(t-1)\alpha']}d\xi \,  \tilde{\mu}_1(\xi)\,[\tilde{M}_2(f_u(\xi))-\tilde{M}_2(f_l(\xi))]\, s^\xi
\end{equation}
where $\tilde{M}_2(X)$ is the primitive of $\tilde{\mu}_2(X)$.\\
For positive $t$'s, the dominant contribution is expected from the upper bound in order to impose the leading Regge trajectory:
\begin{equation}
    M(s,t) \sim s^{\max[1,1+(t-1)\alpha']}
\end{equation}
which is consistent with the positivity of $\xi$ and the large-$s$ limit, at least for $t \geq 1$.\\
In the negative-$t$ region, the integration contour has contributions from both positive and (predominantly) negative $\xi$. If the above behavior for positive $t$ holds and if the weight function is regular, we then obtain:
\begin{equation}
    M(s,t) \sim s
\end{equation}
for negative $t$'s.\\
It is possible to enforce a logarithmic correction to recover results analogous to \eqref{Regge negative t with log}.\\
If $f(\xi)$ has support only in the negative-$\xi$ region and $f(\xi) \sim \xi^{\gamma-1}$, we find
\begin{equation}
    M(s,t) \sim \int^0_{\alpha't} d\xi \, \xi^{\gamma-1} s^\xi \approx \log(s)^{-\gamma}  \sim o(s^0)
\end{equation}
This is the case, for example, in \cite{Veneziano:2017cks}. This latter situation reflects a particular choice of coefficients, which enforces this behavior by suppressing all contributions that would otherwise asymptotically correspond to positive intercepts and zero slopes.\\
The physical explanation of the behaviour in the large-$s$ limit with physical $t$ is that the theory is controlled by an infinite number of LS-amplitudes with a first spin-1 (and a daughter spin-0) resonance at the mass of the $\rho$ meson and an extremely massive spin-2. The corresponding EFT for this model is given by:
\begin{equation}
    M(s,t) \sim \dfrac{ P_0\Big(1+\tfrac{2t}{s}\Big)+ P_1\Big(1+\tfrac{2t}{s}\Big)}{s-m_\rho^2}+ (s \longleftrightarrow t)
\end{equation}
which at large $s$ behaves as
\begin{equation}
    M(s,t) \sim P_1\Big(1+\tfrac{2s}{t}\Big)\xrightarrow{} s
\end{equation}
Finally, let us briefly comment on the extent to which we can trust our approximations. From \eqref{Ansatz} it is clear that the Stirling approximation of each LS-amplitude in the sum is valid if:
\begin{equation}
    \dfrac{\alpha' s}{k_{\max}} \gg 1
\end{equation}
If the sum over $k$ is unbounded, then \eqref{Regge limit 1} is no longer valid for all the amplitudes in the sum, and our approximations cannot be trusted. However, for positive $t$ their contributions are strongly subleading and we can trust our result. If $t$ is negative, the dominant contribution comes from amplitudes with almost flat slopes. Thus, if we identify as $k^* \sim \alpha' s$, our error on \eqref{Regge negative t} will be of order $o(s^{\alpha' t/k^*}) \sim o(s^{t/s})$. Focusing on the correction to the Regge trajectory, we can trust our results up to $s \gg t$ for the leading behavior.\\
With our methods we are not able to recover analytically the transition between Regge behavior and fixed-angle high-energy behavior. In fact, we first took the large-$s$ limit with fixed $t$, and only subsequently the large-$t$ limit. As previously discussed, we cannot trust the Regge behavior of this model if $t$ is of the same order of magnitude as $s$. The result will therefore differ from the large-$s$ limit with fixed $t/s$ ratio, since we neglected the contributions coming from $t$ in the denominator of \eqref{Ansatz}.\\

\subsection{High-Energy Fixed-Angle Limit}\label{High-Energy Fixed-Angle Limit appendix}
In this section we prove the results shown in section \ref{Fixed-Angle High-Energy Limit}.\\
We have two different asymptotic regions, namely one where $-t,s\gg k$, where we can use the Stirling approximation, and one where $k\gg -t,s$ and we can expand at large $k$.

\subsubsection{\texorpdfstring{Finite \(k\): Stirling regime}
                               {Finite k: Stirling regime}}

For fixed \(k\), Stirling gives
\begin{align}
A_{n,k}(s,t)
\sim&
\frac{1}{\sqrt{k}}
\left(\frac{s}{2k}\right)^{
-\frac{-1+k-2n+s}{2k}
}
\left(
\frac{s+t}{2k}
\right)^{
\frac{-2+3k-4n+s+t}{2k}
}
\left(
-\frac{t}{2k}
\right)^{
-\frac{-1+k-2n+t}{2k}
}.
\end{align}

Using
\begin{equation}
t=\frac{s}{2}(z-1),
\end{equation}
one finds
\begin{equation}
A_{n,k}(s,z)
\sim
\frac{\sqrt{s}}{\sqrt{k}}
2^{-s/(2k)}
(1-z)^{\frac{s-sz}{4k}}
(1+z)^{\frac{s+sz}{4k}}
\,
f_{n,k}(z),
\end{equation}
where
\begin{equation}
f_{n,k}(z)
=
(2-2z)^{\frac{1+2n}{2k}}
(1+z)^{-\frac{1+2n}{k}}.
\end{equation}

Equivalently,
\begin{equation}
A_{n,k}(s,z)
\sim
\frac{\sqrt{s}}{\sqrt{k}}
\exp\!\left[
-\frac{s}{k}F(z)
\right]
f_{n,k}(z),
\end{equation}
with
\begin{equation}
F(z)
=
\frac12\log 2
-\frac{1-z}{4}\log(1-z)
-\frac{1+z}{4}\log(1+z).
\end{equation}

For
\begin{equation}
-1<z<1,
\end{equation}
we have
\begin{equation}
F(z)>0,
\end{equation}
hence every fixed-\(k\) sector is exponentially suppressed with a suppression factor:
\begin{equation}
\exp\!\left[-\frac{s}{k}F(z)\right].
\end{equation}

Therefore the dominant region at large \(s\) is
\begin{equation}
k \sim s.
\end{equation}

Using a saddle point approximation, we obtain a contribution of the form:
\begin{equation}
M_{Stirling}(s,t)=c_{n, s}
\exp\!\left[
-F(z)
\right]
f_{n,s}(z),
\end{equation}

\subsubsection{\texorpdfstring{Large \(k\) regime}
                               {Large k regime}}

Expanding directly at large \(k\),
\begin{equation}
A_{n,k}(s,t)
=
\frac{
\Gamma\!\left(
\frac{1+2n-s}{2k}
\right)
\Gamma\!\left(
\frac{1+2n-t}{2k}
\right)
}{
\Gamma\!\left(
-1+\frac{2+4n-s-t}{2k}
\right)
},
\end{equation}
gives
\begin{equation}
A_{n,k}(s,t)
\sim
2k
\left(
\frac{1}{-1-2n+s}
+
\frac{1}{-1-2n+t}
\right).
\end{equation}

At fixed angle,
\begin{equation}
t=\frac{s}{2}(z-1),
\end{equation}
hence
\begin{equation}
A_{n,k}(s,z)
\sim
-\frac{
2k(1+z)
}{
(1-z)s
}.
\end{equation}

This regime is purely power-like.

No exponential suppression survives.

Assume
\begin{equation}
c_{n,k}
\sim
\frac{c_n}{k^\gamma}.
\end{equation}

Then the large-\(k\) tail behaves as
\begin{equation}
M_{\rm tail}(s,z)
\sim
\sum_{k\gtrsim s}
\frac{k^{1-\gamma}}{s}\simeq \frac{\zeta (\gamma -1,s)}{s}.
\end{equation}

Replacing the sum with an integral as in \cite{Veneziano:2017cks},
\begin{equation}
M_{\rm tail}(s,z)
\simeq \frac{\zeta (\gamma -1,s)}{s}
\sim
\frac1s
\int_0^\frac1s
\frac{d \lambda}{\lambda^2}\,
\lambda^{\gamma-1}
\sim
s^{1-\gamma}.
\end{equation}
where we defined $k \equiv \lambda^{-1}$.
\begin{equation}
\gamma>2
\end{equation}
is required for absolute convergence of the \(k\)-sum.

The full asymptotic behavior is therefore
\begin{equation}
M(s,z)
\sim
s^{1-\gamma}.
\end{equation}

\section{Spectra of the Extremal Solutions}\label{Spectra of the Extremal Solutions}
\subsection{\texorpdfstring{$\{\tilde g_{20},\tilde g_{30}\}$-extremal spectra}
                           {{g20,g30}-extremal spectra}}
\begin{figure}[H]
    \centering
    \includegraphics[width=0.75\linewidth]{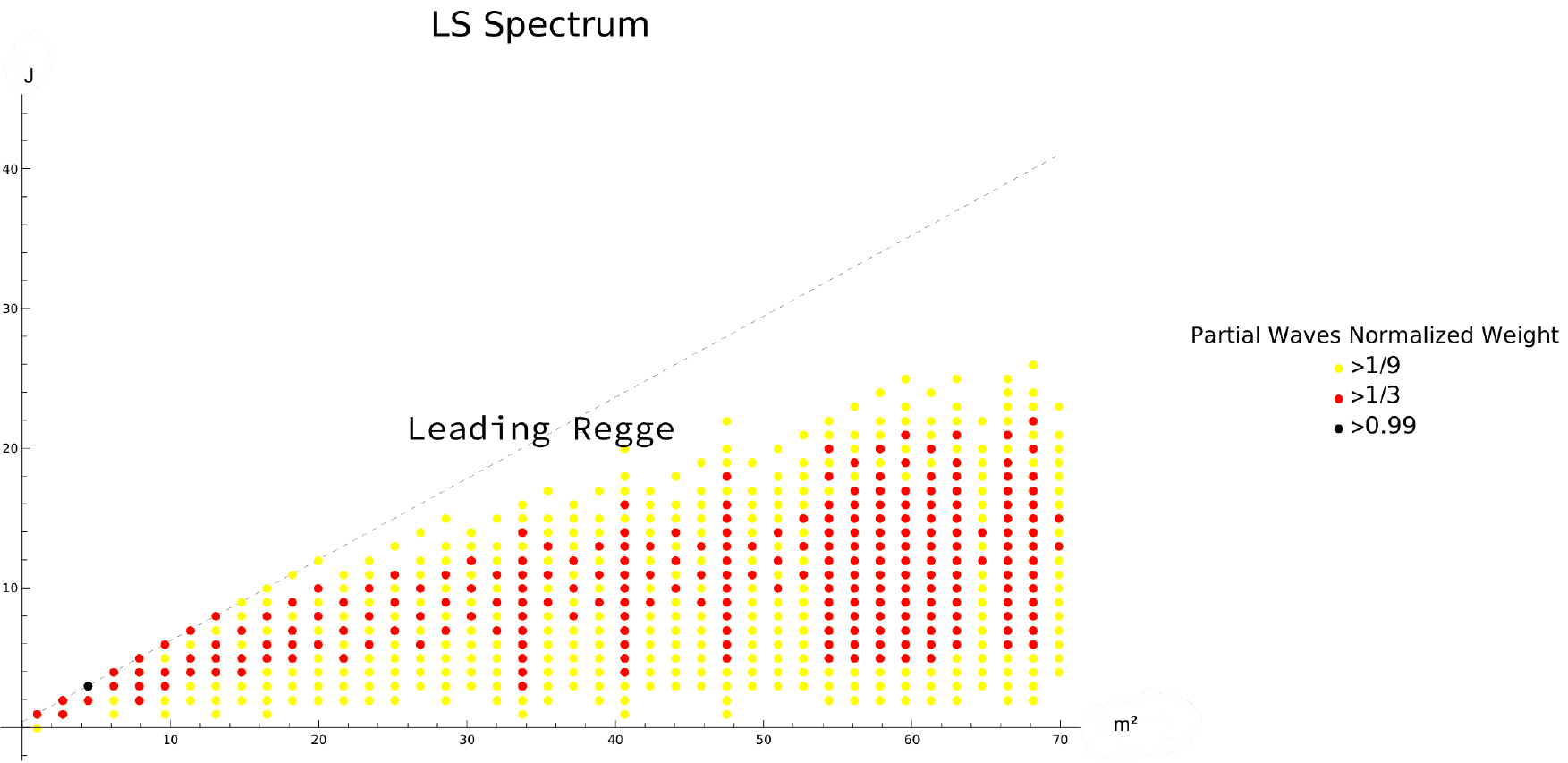}
    \caption{Physical spectrum of the Lovelace–Shapiro amplitude with experimental masses for $m_\rho$ and $m_{f_2}$.}
    \label{fig:Spectrum of the original ls}
\end{figure}

It is possible to numerically study the spectra associated with these extremal solutions. 
We have studied a representative point in the physical-mass case, and we do not expect significant differences in the ideal-mass case.\\
\begin{figure}[H]
    \centering
    \includegraphics[width=0.8\linewidth]{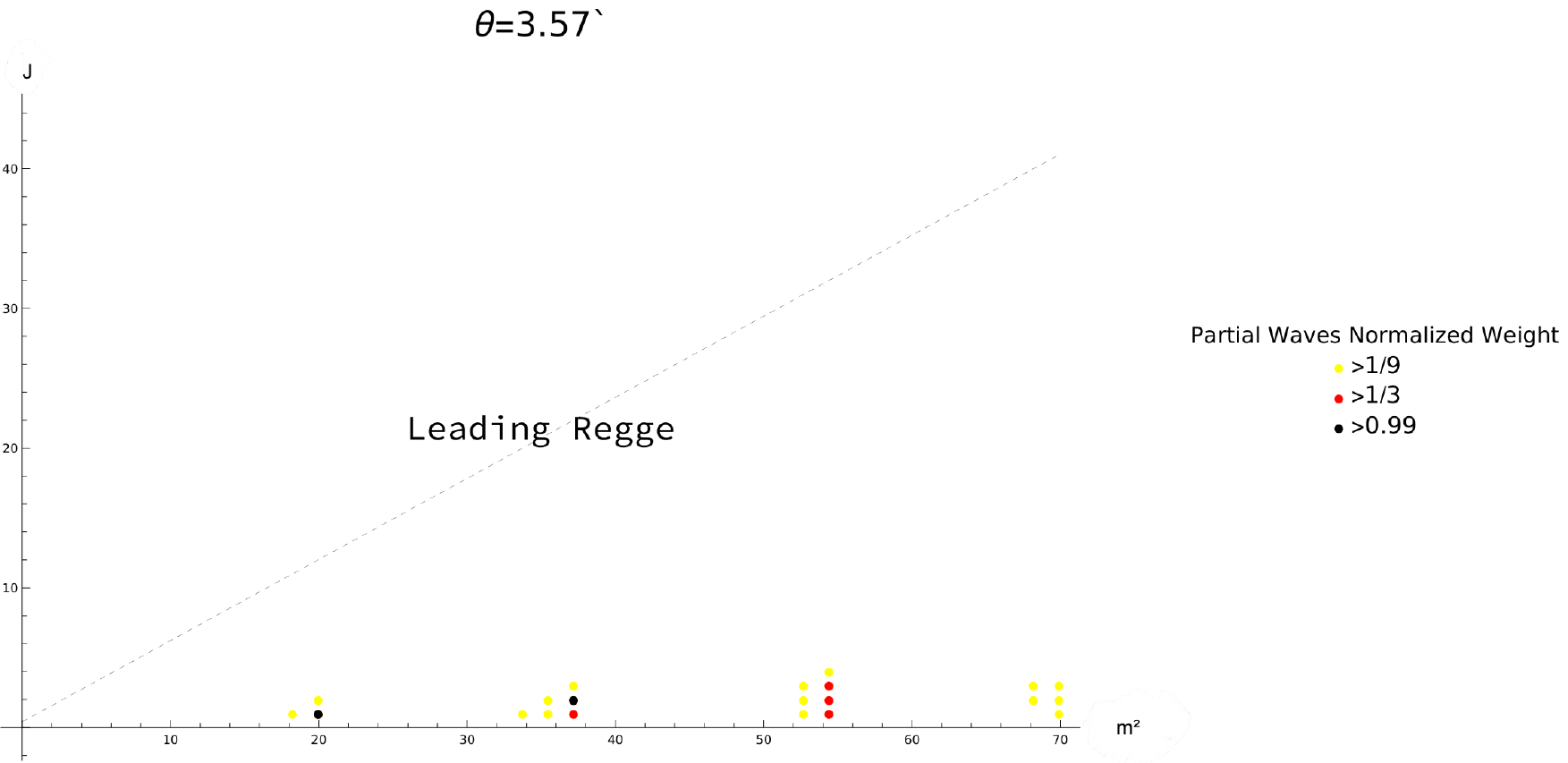}
    \caption{Spectrum of the extremal solution for our ansatz near the region corresponding to pure spin-$1$ exchanges.}
    \label{fig:spin1 exchange spectrum}
\end{figure}
As a consistency check, we examined the spectrum in the region where we would expect a pure spin-1 exchange. We observed the correct structure, apart from some additional heavier particles. Nevertheless, our ansatz successfully cancels most of the couplings between the pion and the other mesons.\\
Another particularly interesting region is the zone near the right kink of our primal bound.
\begin{figure}[H]
    \centering
    \begin{subcaptionbox}{Spectrum below the kink. \label{fig:below}}[0.75\textwidth]
        {\includegraphics[width=\linewidth]{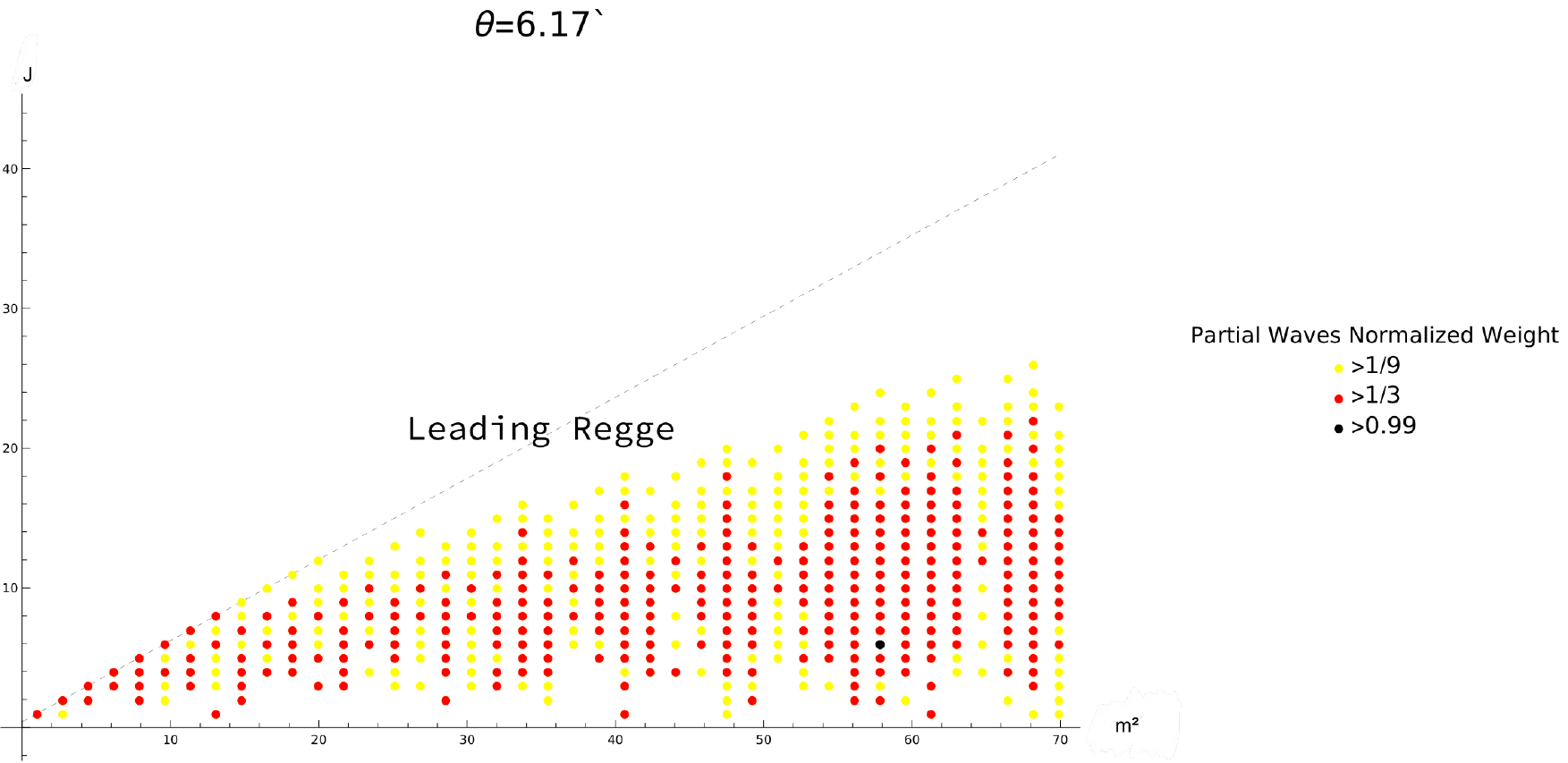}}
    \end{subcaptionbox}
    \vspace{0.3cm}
    \begin{subcaptionbox}{Spectrum near the kink. \label{fig:kink spectrum}}[0.75\textwidth]
        {\includegraphics[width=\linewidth]{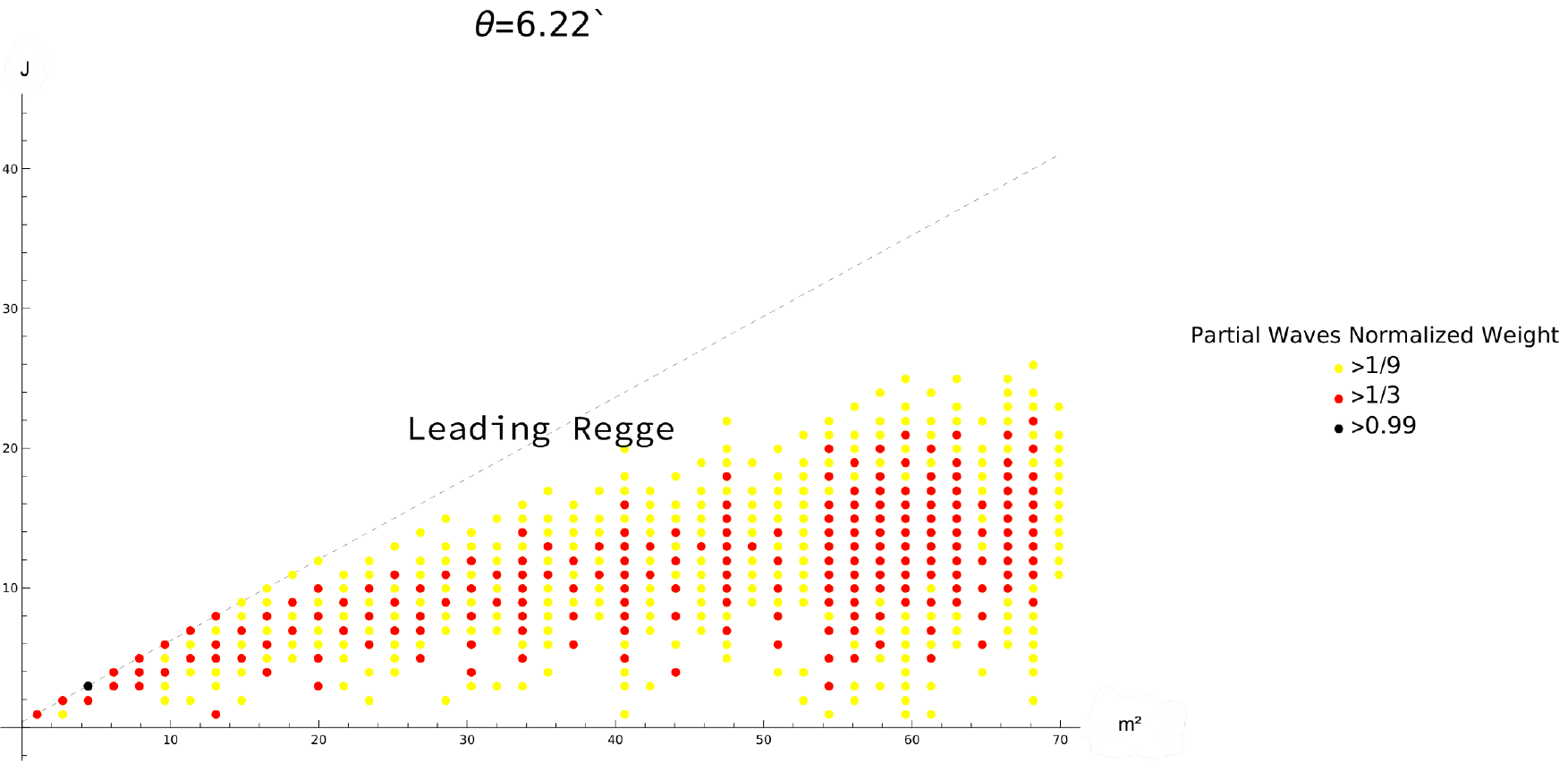}}
    \end{subcaptionbox}
    \vspace{0.3cm}
    \begin{subcaptionbox}{Spectrum above the kink. \label{fig:above}}[0.75\textwidth]
        {\includegraphics[width=\linewidth]{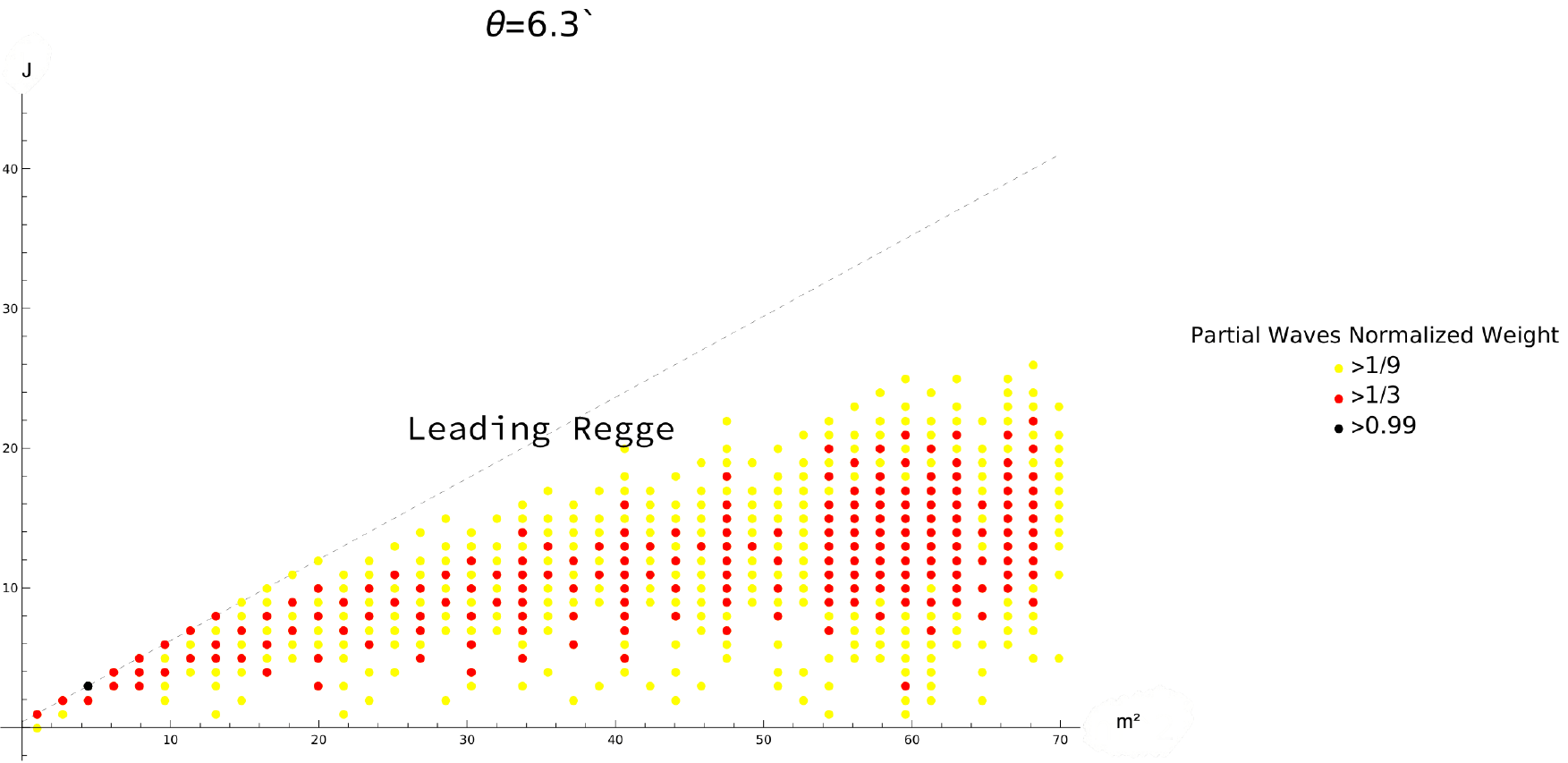}}
    \end{subcaptionbox}
    \caption{Spectrum evolution near the right-kink region.}
    \label{fig:Spectrum Evolution near the kink}
\end{figure}
Approaching the kink along the boundary from below, the dominant coupling shifts from the high-energy resonances to the leading Regge trajectory. The coupling of the spectrum above the kink with the pions is similar to that of the original LS model, shown in figure \ref{fig:Spectrum of the original ls}.\\

\subsection{\texorpdfstring{$g_{\pi\pi f_2}$-extremal spectra}
                           {g pi pi f2-extremal spectra}}
\begin{figure}[H]
    \centering
    \includegraphics[width=0.75\linewidth]{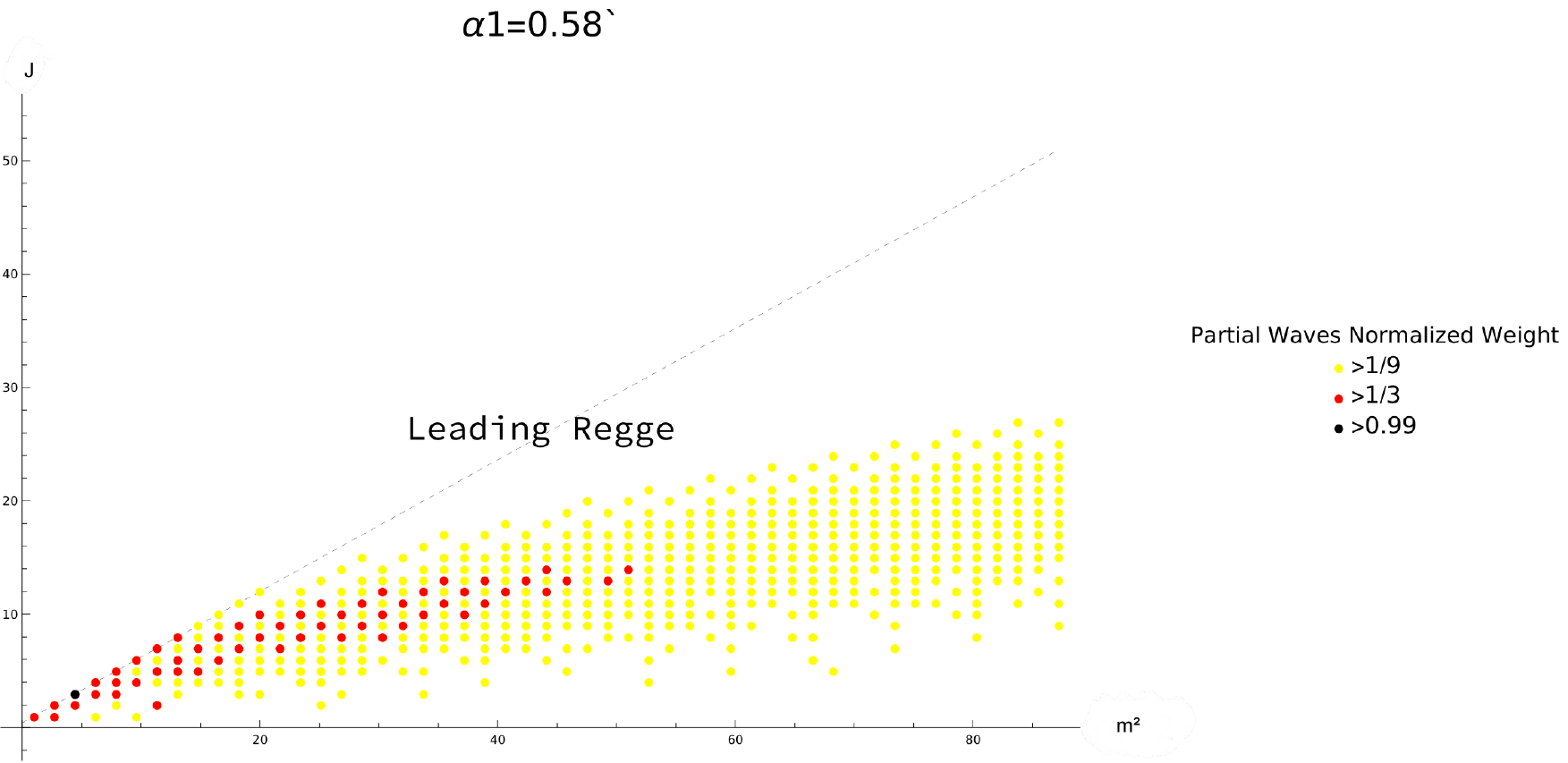}
    \caption{Spectrum of the extremal solutions for the coupling between the $f_2$ meson and the pions.}
    \label{fig:spectrum gf2}
\end{figure}

We have studied the spectrum for $k=6$. We can observe a larger contribution from partial waves in the central region. The main difference is the suppression of higher-mass partial waves, which are more decoupled than the original theory. This is because the extremal value of the $\pi \pi f_2$ coupling is reached through the suppression of the couplings with the heavier mesons.\\
In figure \ref{fig:spectrum gf2} we present the result for $\alpha =0.58$, which is approximately the value corresponding to the physical $\rho$ and $f_2$ masses.

\bibliography{refs.bib}

@book{Collins:1977jy,
    author = "Collins, P. D. B.",
    title = "{An Introduction to Regge Theory and High Energy Physics}",
    doi = "10.1017/9781009403269",
    isbn = "978-1-009-40326-9, 978-1-009-40329-0, 978-1-009-40328-3, 978-0-521-11035-8",
    publisher = "Cambridge University Press",
    series = "Cambridge Monographs on Mathematical Physics",
    month = "7",
    year = "2023"
}

@article{Weinberg:1965nx,
    author = "Weinberg, Steven",
    title = "{Infrared photons and gravitons}",
    doi = "10.1103/PhysRev.140.B516",
    journal = "Phys. Rev.",
    volume = "140",
    pages = "B516--B524",
    year = "1965"
}

@article{Lehmann:1958ita,
    author = "Lehmann, H.",
    title = "{Analytic properties of scattering amplitudes as functions of momentum transfer}",
    doi = "10.1007/bf02859794",
    journal = "Nuovo Cim.",
    volume = "10",
    number = "4",
    pages = "579--589",
    year = "1958"
}

@article{Haring:2022cyf,
    author = {H{\"a}ring, Kelian and Zhiboedov, Alexander},
    title = "{Gravitational Regge bounds}",
    eprint = "2202.08280",
    archivePrefix = "arXiv",
    primaryClass = "hep-th",
    reportNumber = "CERN-TH-2022-016",
    doi = "10.21468/SciPostPhys.16.1.034",
    journal = "SciPost Phys.",
    volume = "16",
    number = "1",
    pages = "034",
    year = "2024"
}

@article{Veneziano:2017cks,
    author = "Veneziano, Gabriele and Yankielowicz, Shimon and Onofri, Enrico",
    title = "{A model for pion-pion scattering in large-N QCD}",
    eprint = "1701.06315",
    archivePrefix = "arXiv",
    primaryClass = "hep-th",
    reportNumber = "CERN-TH-2017-002, TAUP-3013-17",
    doi = "10.1007/JHEP04(2017)151",
    journal = "JHEP",
    volume = "04",
    pages = "151",
    year = "2017"
}

@article{Haring:2023zwu,
    author = {H\"aring, Kelian and Zhiboedov, Alexander},
    title = "{The Stringy S-matrix Bootstrap: Maximal Spin and Superpolynomial Softness}",
    eprint = "2311.13631",
    archivePrefix = "arXiv",
    primaryClass = "hep-th",
    reportNumber = "CERN-TH-2023-214",
    month = "11",
    year = "2023"
}

@article{Correia:2020xtr,
    author = "Correia, Miguel and Sever, Amit and Zhiboedov, Alexander",
    title = "{An analytical toolkit for the S-matrix bootstrap}",
    eprint = "2006.08221",
    archivePrefix = "arXiv",
    primaryClass = "hep-th",
    reportNumber = "CERN-TH-2020-095",
    doi = "10.1007/JHEP03(2021)013",
    journal = "JHEP",
    volume = "03",
    pages = "013",
    year = "2021"
}

@article{Albert:2022oes,
    author = "Albert, Jan and Rastelli, Leonardo",
    title = "{Bootstrapping pions at large N}",
    eprint = "2203.11950",
    archivePrefix = "arXiv",
    primaryClass = "hep-th",
    reportNumber = "YITP-SB-2022-07",
    doi = "10.1007/JHEP08(2022)151",
    journal = "JHEP",
    volume = "08",
    pages = "151",
    year = "2022"
}

@article{Caron-Huot:2020cmc,
    author = "Caron-Huot, Simon and Van Duong, Vincent",
    title = "{Extremal Effective Field Theories}",
    eprint = "2011.02957",
    archivePrefix = "arXiv",
    primaryClass = "hep-th",
    doi = "10.1007/JHEP05(2021)280",
    journal = "JHEP",
    volume = "05",
    pages = "280",
    year = "2021"
}

@article{Caron-Huot:2016icg,
    author = "Caron-Huot, Simon and Komargodski, Zohar and Sever, Amit and Zhiboedov, Alexander",
    title = "{Strings from Massive Higher Spins: The Asymptotic Uniqueness of the Veneziano Amplitude}",
    eprint = "1607.04253",
    archivePrefix = "arXiv",
    primaryClass = "hep-th",
    doi = "10.1007/JHEP10(2017)026",
    journal = "JHEP",
    volume = "10",
    pages = "026",
    year = "2017"
}

@article{Fernandez:2022kzi,
    author = "Fernandez, Clara and Pomarol, Alex and Riva, Francesco and Sciotti, Francesco",
    title = "{Cornering large-N$_{c}$ QCD with positivity bounds}",
    eprint = "2211.12488",
    archivePrefix = "arXiv",
    primaryClass = "hep-th",
    doi = "10.1007/JHEP06(2023)094",
    journal = "JHEP",
    volume = "06",
    pages = "094",
    year = "2023"
}

@article{Tolley:2020gtv,
    author = "Tolley, Andrew J. and Wang, Zi-Yue and Zhou, Shuang-Yong",
    title = "{New positivity bounds from full crossing symmetry}",
    eprint = "2011.02400",
    archivePrefix = "arXiv",
    primaryClass = "hep-th",
    doi = "10.1007/JHEP05(2021)255",
    journal = "JHEP",
    volume = "05",
    pages = "255",
    year = "2021"
}

@article{Bellazzini:2020cot,
    author = "Bellazzini, Brando and Elias Mir{\'o}, Joan and Rattazzi, Riccardo and Riembau, Marc and Riva, Francesco",
    title = "{Positive moments for scattering amplitudes}",
    eprint = "2011.00037",
    archivePrefix = "arXiv",
    primaryClass = "hep-th",
    doi = "10.1103/PhysRevD.104.036006",
    journal = "Phys. Rev. D",
    volume = "104",
    number = "3",
    pages = "036006",
    year = "2021"
}

@article{Arkani-Hamed:2020blm,
    author = "Arkani-Hamed, Nima and Huang, Tzu-Chen and Huang, Yu-tin",
    title = "{The EFT-Hedron}",
    eprint = "2012.15849",
    archivePrefix = "arXiv",
    primaryClass = "hep-th",
    reportNumber = "NCTS-TH/2014, CALT-TH 2020-061",
    doi = "10.1007/JHEP05(2021)259",
    journal = "JHEP",
    volume = "05",
    pages = "259",
    year = "2021"
}

@article{Poland:2018epd,
    author = "Poland, David and Rychkov, Slava and Vichi, Alessandro",
    title = "{The Conformal Bootstrap: Theory, Numerical Techniques, and Applications}",
    eprint = "1805.04405",
    archivePrefix = "arXiv",
    primaryClass = "hep-th",
    doi = "10.1103/RevModPhys.91.015002",
    journal = "Rev. Mod. Phys.",
    volume = "91",
    pages = "015002",
    year = "2019"
}

@article{Rattazzi:2008pe,
    author = "Rattazzi, Riccardo and Rychkov, Vyacheslav S. and Tonni, Erik and Vichi, Alessandro",
    title = "{Bounding scalar operator dimensions in 4D CFT}",
    eprint = "0807.0004",
    archivePrefix = "arXiv",
    primaryClass = "hep-th",
    doi = "10.1088/1126-6708/2008/12/031",
    journal = "JHEP",
    volume = "12",
    pages = "031",
    year = "2008"
}

@article{Li:2023qzs,
    author = "Li, Yue-Zhou",
    title = "{Effective field theory bootstrap, large-N {\ensuremath{\chi}}PT and holographic QCD}",
    eprint = "2310.09698",
    archivePrefix = "arXiv",
    primaryClass = "hep-th",
    doi = "10.1007/JHEP01(2024)072",
    journal = "JHEP",
    volume = "01",
    pages = "072",
    year = "2024"
}

@article{Ma:2023vgc,
    author = "Ma, Teng and Pomarol, Alex and Sciotti, Francesco",
    title = "{Bootstrapping the chiral anomaly at large N$_{c}$}",
    eprint = "2307.04729",
    archivePrefix = "arXiv",
    primaryClass = "hep-th",
    doi = "10.1007/JHEP11(2023)176",
    journal = "JHEP",
    volume = "11",
    pages = "176",
    year = "2023"
}

@article{Regge:1959mz,
    author = "Regge, T.",
    title = "{Introduction to complex orbital momenta}",
    doi = "10.1007/BF02728177",
    journal = "Nuovo Cim.",
    volume = "14",
    pages = "951",
    year = "1959"
}

@article{Lovelace:1968kjy,
    author = "Lovelace, C.",
    title = "{A novel application of regge trajectories}",
    doi = "10.1016/0370-2693(68)90255-4",
    journal = "Phys. Lett. B",
    volume = "28",
    pages = "264--268",
    year = "1968"
}

@article{Shapiro:1969km,
    author = "Shapiro, J. A.",
    title = "{Narrow-resonance model with regge behavior for pi pi scattering}",
    doi = "10.1103/PhysRev.179.1345",
    journal = "Phys. Rev.",
    volume = "179",
    pages = "1345--1353",
    year = "1969"
}

@article{HOOFT1974461,
title = {A planar diagram theory for strong interactions},
journal = {Nuclear Physics B},
volume = {72},
number = {3},
pages = {461-473},
year = {1974},
issn = {0550-3213},
doi = {https://doi.org/10.1016/0550-3213(74)90154-0},
url = {https://www.sciencedirect.com/science/article/pii/0550321374901540},
author = {G.'t Hooft}
}

@article{WITTEN197957,
title = {Baryons in the 1N expansion},
journal = {Nuclear Physics B},
volume = {160},
number = {1},
pages = {57-115},
year = {1979},
issn = {0550-3213},
doi = {https://doi.org/10.1016/0550-3213(79)90232-3},
url = {https://www.sciencedirect.com/science/article/pii/0550321379902323},
author = {Edward Witten}
}

@article{OKUBO1963165,
title = {$\phi$-meson and unitary symmetry model},
journal = {Physics Letters},
volume = {5},
number = {2},
pages = {165-168},
year = {1963},
issn = {0031-9163},
doi = {https://doi.org/10.1016/S0375-9601(63)92548-9},
url = {https://www.sciencedirect.com/science/article/pii/S0375960163925489},
author = {S. Okubo}
}

@article{Polchinski:2001tt,
    author = "Polchinski, Joseph and Strassler, Matthew J.",
    title = "{Hard scattering and gauge / string duality}",
    eprint = "hep-th/0109174",
    archivePrefix = "arXiv",
    reportNumber = "NSF-ITP-01-76, UPR-956-T",
    doi = "10.1103/PhysRevLett.88.031601",
    journal = "Phys. Rev. Lett.",
    volume = "88",
    pages = "031601",
    year = "2002"
}

@article{10.1143/PTPS.37.21,
    author = {Iizuka, Jugoro},
    title = "{A Systematics and Phenomenology of Meson Family*}",
    journal = {Progress of Theoretical Physics Supplement},
    volume = {37-38},
    pages = {21-34},
    year = {1966},
    month = {03},
    issn = {0375-9687},
    doi = {10.1143/PTPS.37.21},
    url = {https://doi.org/10.1143/PTPS.37.21},
    eprint = {https://academic.oup.com/ptps/article-pdf/doi/10.1143/PTPS.37.21/5215468/37-38-21.pdf},
}

@misc{Zweig:570209,
  author       = {Zweig, G.},
  title        = {An SU(3) Model for Strong Interaction Symmetry and its Breaking. Version 2.},
  year         = {1964},
  reportNumber = {CERN-TH-412},
  doi          = {10.17181/CERN-TH-412},
  url          = {https://cds.cern.ch/record/570209},
  note         = {CERN preprint. The archived version differs from the corrected version circulated in 1964.}
}

@article{Collins1984,
  author    = {P. D. B. Collins and P. J. Kearney},
  title     = {Regge theory and QCD in large-angle scattering},
  journal   = {Zeitschrift für Physik C Particles and Fields},
  volume    = {22},
  number    = {3},
  pages     = {277--288},
  year      = {1984},
  month     = sep,
  doi       = {10.1007/BF01575793},
  url       = {https://doi.org/10.1007/BF01575793}
}

@article{Brodsky.31.1153,
  title = {Scaling Laws at Large Transverse Momentum},
  author = {Brodsky, Stanley J. and Farrar, Glennys R.},
  journal = {Phys. Rev. Lett.},
  volume = {31},
  issue = {18},
  pages = {1153--1156},
  numpages = {0},
  year = {1973},
  month = {Oct},
  publisher = {American Physical Society},
  doi = {10.1103/PhysRevLett.31.1153},
  url = {https://link.aps.org/doi/10.1103/PhysRevLett.31.1153}
}

@article{Albert:2023jtd,
    author = "Albert, Jan and Rastelli, Leonardo",
    title = "{Bootstrapping pions at large N. Part II. Background gauge fields and the chiral anomaly}",
    eprint = "2307.01246",
    archivePrefix = "arXiv",
    primaryClass = "hep-th",
    reportNumber = "YITP-SB-2023-15",
    doi = "10.1007/JHEP09(2024)039",
    journal = "JHEP",
    volume = "09",
    pages = "039",
    year = "2024"
}

@article{Albert:2023seb,
    author = "Albert, Jan and Henriksson, Johan and Rastelli, Leonardo and Vichi, Alessandro",
    title = "{Bootstrapping mesons at large N: Regge trajectory from spin-two maximization}",
    eprint = "2312.15013",
    archivePrefix = "arXiv",
    primaryClass = "hep-th",
    reportNumber = "YITP-SB-2023-41",
    doi = "10.1007/JHEP09(2024)172",
    journal = "JHEP",
    volume = "09",
    pages = "172",
    year = "2024"
}

@article{McPeak:2023wmq,
    author = "McPeak, Brian and Venuti, Marco and Vichi, Alessandro",
    title = "{Adding subtractions: comparing the impact of different Regge behaviors}",
    eprint = "2310.06888",
    archivePrefix = "arXiv",
    primaryClass = "hep-th",
    month = "10",
    year = "2023"
}

@article{Guerrieri:2020bto,
    author = "Guerrieri, Andrea L. and Penedones, Joao and Vieira, Pedro",
    title = "{S-matrix bootstrap for effective field theories: massless pions}",
    eprint = "2011.02802",
    archivePrefix = "arXiv",
    primaryClass = "hep-th",
    doi = "10.1007/JHEP06(2021)088",
    journal = "JHEP",
    volume = "06",
    pages = "088",
    year = "2021"
}

@article{Guerrieri:2018uew,
    author = "Guerrieri, Andrea L. and Penedones, Joao and Vieira, Pedro",
    title = "{Bootstrapping QCD Using Pion Scattering Amplitudes}",
    eprint = "1810.12849",
    archivePrefix = "arXiv",
    primaryClass = "hep-th",
    doi = "10.1103/PhysRevLett.122.241604",
    journal = "Phys. Rev. Lett.",
    volume = "122",
    number = "24",
    pages = "241604",
    year = "2019"
}

@article{Mandelstam:1958xc,
    author = "Mandelstam, S.",
    title = "{Determination of the pion - nucleon scattering amplitude from dispersion relations and unitarity. General theory}",
    doi = "10.1103/PhysRev.112.1344",
    journal = "Phys. Rev.",
    volume = "112",
    pages = "1344--1360",
    year = "1958"
}

@article{Olive1962,
  author    = {D. I. Olive},
  title     = {Unitarity and the evaluation of discontinuities},
  journal   = {Il Nuovo Cimento (1955-1965)},
  volume    = {26},
  number    = {1},
  pages     = {73--102},
  year      = {1962},
  month     = oct,
  doi       = {10.1007/BF02754344},
  url       = {https://doi.org/10.1007/BF02754344},
  issn      = {1827-6121}
}

@article{Sommer:1970mr,
    author = "Sommer, G.",
    title = "{Present state of rigorous analytic properties of scattering amplitudes}",
    doi = "10.1002/prop.19700181102",
    journal = "Fortsch. Phys.",
    volume = "18",
    pages = "577--688",
    year = "1970"
}

@article{Bros:1964iho,
    author = "Bros, J. and Epstein, H. and Glaser, Vladimir Jurko",
    title = "{Some rigorous analyticity properties of the four-point function in momentum space}",
    reportNumber = "CERN-TH-358",
    doi = "10.1007/BF02733596",
    journal = "Nuovo Cim.",
    volume = "31",
    pages = "1265--1302",
    year = "1964"
}

@article{PhysRev.137.B1022,
  title = {Consistency Conditions on the Strong Interactions Implied by a Partially Conserved Axial-Vector Current},
  author = {Adler, Stephen L.},
  journal = {Phys. Rev.},
  volume = {137},
  issue = {4B},
  pages = {B1022--B1033},
  numpages = {0},
  year = {1965},
  month = {Feb},
  publisher = {American Physical Society},
  doi = {10.1103/PhysRev.137.B1022},
  url = {https://link.aps.org/doi/10.1103/PhysRev.137.B1022}
}

@book{Weinberg:1996kr,
    author = "Weinberg, Steven",
    title = "{The quantum theory of fields. Vol. 2: Modern applications}",
    doi = "10.1017/CBO9781139644174",
    isbn = "978-1-139-63247-8, 978-0-521-67054-8, 978-0-521-55002-4",
    publisher = "Cambridge University Press",
    month = "8",
    year = "2013"
}

@article{FigueroaTourkine2022,
  author        = {Figueroa, Felipe and Tourkine, Piotr},
  title         = {On the unitarity and low energy expansion of the Coon amplitude},
  year          = {2022},
  eprint        = {2201.12331},
  archivePrefix = {arXiv},
  primaryClass  = {hep-th},
  doi           = {10.1103/PhysRevLett.129.121602}
}

@article{MaldacenaRemmen2022,
  author        = {Maldacena, Juan and Remmen, Grant N.},
  title         = {Accumulation-Point Amplitudes in String Theory},
  journal       = {JHEP},
  volume        = {08},
  pages         = {152},
  year          = {2022},
  eprint        = {2207.06426},
  archivePrefix = {arXiv},
  primaryClass  = {hep-th},
  doi           = {10.1007/JHEP08(2022)152}
}

@article{GumusLeflotTourkineZhiboedov2024,
  author        = {Gumus, Mehmet Asim and Leflot, Damien and Tourkine, Piotr and Zhiboedov, Alexander},
  title         = {The S-matrix bootstrap with neural optimizers I: zero double discontinuity},
  year          = {2024},
  eprint        = {2412.09610},
  archivePrefix = {arXiv},
  primaryClass  = {hep-th}
}

@article{GumusLeflotTourkineZhiboedov2026,
  author        = {Gumus, Mehmet Asim and Leflot, Damien and Tourkine, Piotr and Zhiboedov, Alexander},
  title         = {Neural S-matrix bootstrap II: solvable 4d amplitudes with particle production},
  year          = {2026},
  eprint        = {2601.22145},
  archivePrefix = {arXiv},
  primaryClass  = {hep-th}
}

@article{Ghosh:2026xnp,
    author = "Ghosh, Kausik and Kumar, Sidhaarth and Niarchos, Vasilis and Stergiou, Andreas",
    title = "{Neural Spectral Bias and Conformal Correlators I: Introduction and Applications}",
    eprint = "2604.18686",
    archivePrefix = "arXiv",
    primaryClass = "hep-th",
    reportNumber = "ITCP-2026-5, CCTP-2-26-5",
    month = "4",
    year = "2026"
}

@article{Ghosh:2026jbw,
    author = "Ghosh, Kausik and Kumar, Sidhaarth and Stergiou, Andreas and Niarchos, Vasilis",
    title = "{Neural Networks Reveal a Universal Bias in Conformal Correlators}",
    eprint = "2604.18673",
    archivePrefix = "arXiv",
    primaryClass = "hep-th",
    reportNumber = "ITCP-2026-4, CCTP-2026-4",
    month = "4",
    year = "2026"
}

@article{Benjamin:2026lbj,
    author = "Benjamin, Nathan and Fitzpatrick, A. Liam and Li, Wei and Thaler, Jesse",
    title = "{Descending into the Modular Bootstrap}",
    eprint = "2604.01275",
    archivePrefix = "arXiv",
    primaryClass = "hep-th",
    reportNumber = "MIT-CTP/6023",
    month = "4",
    year = "2026"
}

\end{document}